\documentclass[journal,11pt,draftcls,onecolumn]{IEEEtran}

\usepackage{graphicx, amsmath, amsthm, amssymb, amsfonts, cancel}
\usepackage{algorithm, algorithmic, ifsym, subfigure}

\def\Cov{{\rm Cov}}
\def\tr{{\rm tr}}
\def\E{ \mathbb{E} }
\def\P{ \mathbb{P} }
\def\bA{ {\mathbf{A}} }
\def\bB{ {\mathbf{B}} } 
\def\bC{ {\mathbf{C}} }

\def\bG{ {\mathbf{G}} } 
 
\def\bI{ {\mathbf{I}} } 

\def\bK{ {\mathbf{K}} } 

\def\bM{ {\mathbf{M}} } 
\def\bN{ {\mathbf{N}} }

\def\bR{ {\mathbf{R}} } 
\def\bS{ {\mathbf{S}} }
\def\bT{ {\mathbf{T}} } 
\def\bU{ {\mathbf{U}} }
 
\def\bW{ {\mathbf{W}} }
\def\bX{ {\mathbf{X}} } 
\def\bY{ {\mathbf{Y}} } 
\def\bZ{ {\mathbf{Z}} } 

\def\bTheta{ {\mathbf{\Theta}} } 
\def\bSigma{ {\mathbf{\Sigma}} } 
\def\bDelta{ {\mathbf{\Delta}} }

\def\nn{{ \parallel   }}
\def\RR{{ \mathbb{R}  }}

\def\NN{{ \mathbb{N}  }}
\def\vec{{ \text{vec} }}
\def\tr{{ \text{tr}   }}

\def\bx{{ \mathbf{x}  }}

\def\bz{{ \mathbf{z}  }}

\def\card{{ \text{card} }}

\newtheorem{example}{Example}
\newtheorem{theorem}{Theorem}
\newtheorem{lemma}{Lemma}
\newtheorem{remark}{Remark}
\newtheorem{corollary}{Corollary}
\newtheorem{proposition}{Proposition}
\newtheorem{assumption}{Assumption}
\newtheorem{definition}{Definition}

\newtheorem{property}{Property}

\begin{document}
\title{Convergence Properties of Kronecker Graphical Lasso Algorithms}
\author{Theodoros Tsiligkaridis *, \textit{Student Member, IEEE}, Alfred O. Hero III, \textit{Fellow, IEEE}, Shuheng Zhou, \textit{Member, IEEE}}

\maketitle

\begin{abstract}
This report presents a thorough convergence analysis of Kronecker graphical lasso (KGLasso) algorithms for estimating the covariance of an i.i.d. Gaussian random sample under a sparse Kronecker-product covariance model. The KGlasso model, originally called the transposable regularized covariance model by Allen {\it et al}  \cite{AllenTib10}, implements a pair of $\ell_1$ penalties on each Kronecker factor to enforce sparsity in the covariance estimator.   The KGlasso algorithm generalizes Glasso, introduced by Yuan and Lin \cite{YL07} and Banerjee {\it et al} \cite{ModelSel}, to estimate covariances having Kronecker product form. It  also generalizes  the unpenalized ML flip-flop (FF) algorithm of Dutilleul \cite{Dutilleul} and Werner {\it et al} \cite{EstCovMatKron} to estimation of sparse Kronecker factors. We establish that the KGlasso iterates converge pointwise to a local maximum of the penalized likelihood function. We derive high dimensional rates of convergence to the true covariance as both the number of samples and the number of variables go to infinity. Our results establish that KGlasso has significantly faster asymptotic convergence than FF and Glasso. Our results establish that KGlasso has significantly faster asymptotic convergence than FF and Glasso. Simulations are presented that validate the results of our analysis.  For example, for a sparse $10,000 \times 10,000$ covariance matrix equal to the Kronecker product of two $100 \times 100$ matrices, the root mean squared error of the inverse covariance estimate using FF is 3.5 times larger than that obtainable using KGlasso.
\end{abstract}


\let\thefootnote\relax\footnote{The research reported in this paper was supported in part by ARO grant W911NF-11-1-0391. Preliminary results in this paper have appeared at the 2012 IEEE International Conference on Acoustics, Speech, and Signal Processing and 2012 IEEE Statistical Signal Processing Workshop.

T. Tsiligkaridis and A. O. Hero, III, are with the Department of Electrical Engineering and Computer Science, University of Michigan, Ann Arbor, MI 48109 USA (e-mail: ttsili@umich.edu, hero@umich.edu).

S. Zhou is with the Department of Statistics, University of Michigan, Ann Arbor, MI 48109 USA (e-mail: shuhengz@umich.edu).
}

\section{Introduction} \label{sec: intro}
Covariance estimation is a problem of great interest in many different disciplines, including machine learning, signal processing, economics and bioinformatics. In many applications the number of variables is very large, e.g., in the tens or hundreds of thousands, leading to a number of covariance parameters that greatly exceeds the number of observations.  To address this problem constraints are frequently imposed on the covariance to reduce the number of parameters in the model. For example, the Glasso model of  Yuan and Lin \cite{YL07} and Banerjee {\it et al} \cite{ModelSel}  imposes sparsity constraints on the covariance. The Kronecker product  model of Dutilleul \cite{Dutilleul} and Werner {\it et al} \cite{EstCovMatKron} assumes that the covariance can be represented as the Kronecker product  of two lower dimensional covariance matrices. The transposable regularized covariance model of Allen {\it et al}  \cite{AllenTib10} imposes a combination of sparsity and Kronecker product form on the covariance. When there is no missing data, an extension of the alternating optimization algorithm of  \cite{Dutilleul, EstCovMatKron}, called the flip flop (FF) algorithm, can be applied to estimate the parameters of this combined sparse and Kronecker product model.  In this report we call this algorithm the Kronecker Glasso (KGlasso) and  we thoroughly analyze convergence of the algorithm in the high dimensional setting.

As in \cite{EstCovMatKron} we assume that there are $pf$ variables whose covariance $\mathbf{\Sigma}_0$ has the separable positive definite Kronecker product representation:
\begin{equation} \label{factorize}
	\mathbf{\Sigma}_{0} = \mathbf{A}_0 \otimes \mathbf{B}_0
\end{equation}
where $\bA_0$ is a $p\times p$ positive definite matrix and $\bB_0$ is an $f \times f$ positive definite matrix.  This model (\ref{factorize}) is relevant to channel modeling for MIMO wireless communications, where $\bA_0$ is a transmit covariance matrix and $\bB_0$ is a receive covariance matrix \cite{MIMOWerner}. The model is also relevant to other transposable models arising in recommendation systems like NetFlix and in gene expression analysis \cite{AllenTib10}.

The Kronecker product Gaussian graphical model has been known for a long time as the matrix normal distribution in the statistics community \cite{Dawid, Dutilleul, GuptaNagar99}. Various properties of the matrix variate normal distribution have been studied in \cite{GuptaNagar99}. Let us rewrite the problem into matrix form. Consider a $p\times f$ random matrix $\bZ$ that follows a matrix normal distribution-i.e. $\bZ \sim N_{p,f}(\mathbf{0}; \bA_0,\bB_0)$ \cite{GuptaNagar99}. Then, $\bB_0$ is the row covariance matrix and $\bA_0$ is the column covariance matrix-i.e., $\bZ_{:,k} \sim N(\mathbf{0},[\bB_0]_{k,k} \bA_0)$ and $\bZ_{i,:} \sim N(\mathbf{0},[\bA_0]_{i,i} \bB_0)$ \footnote{Here, $\bZ_{i,:}$ is the $i$th row and $\bZ_{:,k}$ is the $k$th column of the matrix $\bZ$. For concreteness, assume $\bz=[\bz_1^T,\dots,\bz_p^T]^T \sim N(\mathbf{0},\bSigma_0)$. Then, $\bZ=[\bz_1,\dots,\bz_p]^T$ is the $p\times f$ data matrix with row covariance $\bB_0$ and column covariance $\bA_0$. }. This model further finds applications in geostatistics \cite{Cressie93} and genomics \cite{YinLi2012}. Further applications of matrix-variate normal models include collaborative filtering \cite{YuICML09}, multi-task learning \cite{BonillaNIPS08} and face recognition \cite{ZhangNIPS10}. The Kronecker factorization (\ref{factorize}) can easily be generalized to the $k$-fold case, where $\mathbf{\Sigma}_0=\mathbf{A}_1\otimes \mathbf{A}_2 \otimes \dots \otimes \mathbf{A}_k$. 

Under the assumption that the measurements are multivariate Gaussian with covariance having the Kronecker product form (\ref{factorize}), the maximum likelihood (ML) estimator can be formulated \cite{LuZimmerman2}. While the ML estimator has no known closed-form solution, an approximation to the solution can be iteratively computed via an alternating algorithm: the flip-flop (FF) algorithm \cite{LuZimmerman2, EstCovMatKron}.  As compared to the standard saturated (unstructured) covariance model, the number of unknown parameters in (\ref{factorize}) is reduced from order $\Theta(p^2 f^2)$ to order $\Theta(p^2)+\Theta(f^2)$.
This results in a significant reduction in the mean squared error (MSE) and the computational complexity of the maximum likelihood (ML) covariance estimator. This report establishes that further reductions MSE are achievable when the Kronecker matrix factors are known to have sparse inverses, i.e., the measurements obey a sparse Kronecker structured Gaussian graphical model.

The graphical lasso (Glasso) estimator was originally proposed in \cite{YL07, ModelSel} for estimating a sparse inverse covariance, also called the precision matrix, under an i.i.d. Gaussian observation model. An algorithm for efficiently solving the nonsmooth optimization problem that arises in the Glasso estimator, based on ideas from \cite{ModelSel}, was proposed in \cite{Glasso}. Glasso has been applied to the time-varying coefficients setting in Zhou {\it et al} \cite{TimeVaryingGraphs} using the kernel estimator for covariances at a target time. Rothman {\it et al} \cite{Rothman} derived high dimensional convergence rates for a slight variant of Glasso, i.e., only the off-diagonal entries of the estimated precision matrix were penalized using an $\ell_1$-penalty. The high dimensional convergence rate of Glasso was established by Ravikumar {\it et al}  \cite{RWRY08}. This report extends their analysis to the case that the covariance has Kronecker structure (\ref{factorize}), showing that significantly higher rates of convergence are achievable.

The main contribution is the derivation of the high-dimensional MSE convergence rates for KGlasso as $n$, $p$ and $f$ go to infinity.
 When both Kronecker factors are sparse, it is shown that KGlasso \textit{strictly} outperforms FF and Glasso in terms of MSE convergence rate. More specifically, we show KGlasso achieves a convergence rate of $O_P\left(\frac{(p+f)\log \max(p,f,n)}{n}\right)$ and FF achieves a rate of $O_P\left(\frac{(p^2+f^2)\log \max(p,f,n)}{n}\right)$ as $n\to\infty$, while it is known \cite{Rothman, TimeVaryingGraphs} that Glasso achieves a rate of $O_P\left(\frac{(pf+s)\log \max(p,f,n)}{n}\right)$, where $s$ denotes the number of off-diagonal nonzero elements in the true precision matrix $\bTheta_0$. Simulations show that the performance improvements predicted by the high-dimensional analysis continue to hold for small sample size and moderate matrix dimension. For the example studied in Sec. \ref{sec: simulations} the empirical MSE of KGlasso is significantly lower than that of Glasso and FF for $p=f=100$ over the range of $n$  from $10$ to $100$.

The starting point for the MSE convergence analysis is the large-sample analysis of the FF algorithm (Thm. 1 in \cite{EstCovMatKron}). The KGlasso convergence proof uses a large deviation inequality  that shows that the dimension of one estimated Kronecker factor, say $\bA$, acts as a multiplier on the number of independent samples when performing inference on the other factor $\bB$. This result is then used to obtain optimal MSE rates in terms of Frobenius norm error between the KGlasso estimated matrix and the ground truth. The asymptotic MSE convergence analysis is useful since it can be used to guide the selection of sparsity regularization parameters and to determine minimum sample size requirements.

An anonymous reviewer alerted the authors to the related work of Yin and Li \cite{YinLi2012},  published after submission of this paper for publication. Yin and Li  obtain high-dimensional MSE bounds for the same matrix normal estimation problem considered here.  However, our MSE bounds are tighter than the bounds given in Yin and Li. In particular, neglecting terms of order $\log(pf)$, our bounds are of  order $p+f$ as compared to Yin and Li's bounds of order $pf$, which is significantly weaker for large $p,f$. We obtain improved bounds due to the use of a tighter concentration inequality, established in Lemma \ref{lemma: large_dev_optimal}.

\subsection{Outline}
The outline of the report is as follows. Section \ref{sec: notation} introduces the notation that will be used throughout the report. In Section \ref{sec: glasso}, the graphical lasso framework is introduced. Section \ref{sec: algorithm} uses this framework to introduce the KGlasso algorithm. Section \ref{sec: convergence_analysis_KGL} shows convergence of KGlasso and characterizes its limit points. The high dimensional MSE convergence rate derivation for the FF algorithm is included in Section \ref{sec: FF_rate}. Section \ref{sec: consistency} presents a high-dimensional MSE rate result that is used to establish the superiority of KGlasso as compared to FF and standard Glasso, under the sparse Kronecker product representation (\ref{factorize}). Section \ref{sec: simulations} presents simulations that empirically validate the theoretical convergence rates obtained in Section \ref{sec: consistency}.

\section{Notation} \label{sec: notation}
For a square matrix $\mathbf{M}$, define $|\mathbf{M}|_1=\nn \vec(\mathbf{M}) \nn_1$ and $|\mathbf{M}|_\infty=\nn \vec(\mathbf{M}) \nn_\infty$, where $\vec(\mathbf{M})$ denotes the vectorized form of $\mathbf{M}$ (concatenation of columns into a vector). $\nn \mathbf{M} \nn_2$ is the spectral norm of $\bM$. $\mathbf{M}_{i,j}$ and $[\mathbf{M}]_{i,j}$ are the $(i,j)$th element of $\mathbf{M}$. Let the inverse transformation (from a vector to a matrix) be defined as: $\vec^{-1}(\mathbf{x})=\mathbf{X}$, where $\mathbf{x}=\vec(\mathbf{X})$. Define the $pf\times pf$ permutation operator $\bK_{p,f}$ such that $\bK_{p,f} \vec(\bN) = \vec(\bN^T)$ for any $p\times f$ matrix $\bN$. For a symmetric matrix $\mathbf{M}$, $\lambda(\mathbf{M})$ will denote the vector of real eigenvalues of $\mathbf{M}$ and define $\lambda_{max}(\mathbf{M})=\nn\bM \nn_2=\max{\lambda_i(\mathbf{M})}$ for p.d. symmetric matrix, and $\lambda_{min}(\mathbf{M}) = \min{\lambda_i(\mathbf{M})}$. Define the sparsity parameter associated with $\bM$ as $s_{M}=\card(\{(i_1,i_2): [\bM]_{i_1,i_2}\neq 0, i_1\neq i_2 \})$. Let $\kappa(\bM):=\frac{\lambda_{max}(\bM)}{\lambda_{min}(\bM)}$ denote the condition number of a symmetric matrix $\bM$.

For a matrix $\bM$ of size $pf\times pf$, let $\{\bM(i,j)\}_{i,j=1}^p$ denote its $f\times f$ block submatrices, where each block submatrix is $\bM(i,j)=[\bM]_{(i-1)f+1:if,(j-1)f+1:jf}$. Also let $\{\overline{\bM}(k,l)\}_{k,l=1}^f$ denote the $p\times p$ block submatrices of the permuted matrix $\overline{\bM}=\bK_{p,f}^T \bM \bK_{p,f}$.

Define the set of symmetric matrices $S^p = \{\mathbf{A}\in \RR^{p\times p}: \mathbf{A}=\mathbf{A}^T\}$, the set of symmetric positive semidefinite (psd) matrices $S_{+}^p=\{\mathbf{A} \in \RR^{p\times p}: \mathbf{A}=\mathbf{A}^T, \mathbf{z}^T \mathbf{A} \mathbf{z} \geq 0, \forall \mathbf{z} \in \RR^p \}$, and the set of symmetric positive definite (pd) matrices $S_{++}^p = \{\mathbf{A}\in \RR^{p \times p} : \mathbf{A}=\mathbf{A}^T, \mathbf{z}^T \mathbf{A} \mathbf{z}>0, \forall \mathbf{z}\neq 0\}$. $\bI_d$ is a $d\times d$ identity matrix. It can be shown that $S_{++}^p$ is a convex set, but is not closed \cite{ConvexOpt}. Note that $S_{++}^p$ is simply the interior of the closed convex cone $S_{+}^p$.

Statistical convergence rates will be denoted by the $O_P(\cdot)$ notation, which is defined as follows. Consider a sequence of real random variables $\{X_n\}_{n \in \NN}$ defined on a probability space $(\Omega,\mathcal{F},P)$ and a deterministic (positive) sequence of reals $\{b_n\}_{n\in \NN}$. By $X_n=O_P(1)$ is meant: $\sup_{n\in \NN}{\P(|X_n|>K)} \to 0$ as $K\to\infty$. 
The notation $X_n=O_P(b_n)$ is equivalent to $\frac{X_n}{b_n}=O_P(1)$. By $X_n=o_p(1)$ is meant $\P(|X_n|>\epsilon) \to 0$ as $n\to\infty$ for any $\epsilon>0$. By $\lambda_n \asymp b_n$ is meant $c_1 \leq \frac{\lambda_n}{b_n} \leq c_2$ for all $n$, where $c_1,c_2>0$ are absolute constants.

\section{Graphical Lasso Framework} \label{sec: glasso}
For simplicity, we assume the number of Kronecker components is $k=2$. Available are $n$ i.i.d. multivariate Gaussian observations $\{\bz_t\}_{t=1}^{n}$, where $\bz_t\in \RR^{pf}$, having zero-mean and covariance equal to $\mathbf{\Sigma} = \mathbf{A}_0 \otimes \mathbf{B}_0$. Then, the log-likelihood is proportional to:
\begin{equation} \label{log_likelihood}
	l(\mathbf{\Sigma}) := \log\det(\mathbf{\Sigma}^{-1}) - \tr(\mathbf{\Sigma}^{-1}\hat{\mathbf{S}}_n),
\end{equation}
where $\mathbf{\Sigma}$ is the positive definite covariance matrix and $\hat{\mathbf{S}}_n=\frac{1}{n}\sum_{t=1}^{n}{\bz_t \bz_t^T}$ is the sample covariance matrix. Recent work \cite{ModelSel, Glasso} has considered $\ell_1$-penalized maximum likelihood estimators for the saturated model where $\bSigma$ belongs to the unrestricted cone of positive definite matrices. These estimators are known as graphical lasso (Glasso) estimators and are the solution to the $\ell_1$-penalized minimization problem:
\begin{equation} \label{opt_prob}
	\hat{\mathbf{\Sigma}}_n \in \arg \min_{\mathbf{\Sigma} \in S_{++}^p} { \{ -l(\mathbf{\Sigma}) + \lambda |\mathbf{\Sigma}^{-1}|_1 \} },
\end{equation}
where $\lambda \geq 0$ is a regularization parameter. If $\lambda>0$ and $\hat{\mathbf{S}}_n$ is positive definite, then $\hat{\mathbf{\Sigma}}_n$ in (\ref{opt_prob}) is the unique minimizer.

A fast iterative algorithm, based on a block coordinate descent approach, exhibiting a computational complexity $\mathcal{O}((pf)^4)$, was developed in \cite{Glasso} to solve the convex program (\ref{opt_prob}). Under the assumption $\lambda \asymp \sqrt{\frac{\log(pf)}{n}}$ solution of (\ref{opt_prob}) was shown to have high dimensional convergence rate \cite{Rothman}:
\begin{equation} \label{Glasso_rate}
 \nn \bG(\hat{\bS}_n, \lambda)-\bTheta_0 \nn_F = O_P\left( \sqrt{\frac{(pf+s) \log(pf)}{n}} \right)
\end{equation}
where $s$ is an upper bound on the number of non-zero off-diagonal elements of $\bTheta_0$. When $s=O(pf)$, this rate is better than the non-regularized sample covariance estimator:
\begin{equation} \label{naive_SCM_rate}
	\nn \hat{\bS}_n - \bSigma_0 \nn_F = O_P\left( \sqrt{\frac{p^2f^2}{n}} \right).
\end{equation}

\section{Kronecker Graphical Lasso} \label{sec: algorithm}
Let $\mathbf{\Sigma}_0:=\mathbf{A}_0 \otimes \mathbf{B}_0$ denote the true covariance matrix, where $\mathbf{A}_0:=\mathbf{X}_0^{-1}$ and $\mathbf{B}_0=\mathbf{Y}_0^{-1}$ are the true Kronecker factors. Let $\mathbf{A}_{init}$ denote the initial guess of $\mathbf{A}_0=\mathbf{X}_0^{-1}$.

Define $J(\bX,\bY)$ as the negative log-likelihood
\begin{align}
	J(\bX,\bY) &= \tr((\bX \otimes \bY)\hat{\bS}_n) - f \log\det(\bX) \nonumber \\
		& \quad -p \log\det(\bY) \label{J_func_2}
\end{align}
Although the objective (\ref{J_func_2}) is not jointly convex in $(\bX, \bY)$, it is biconvex. This motivates the flip-flop algorithm \cite{Dutilleul, EstCovMatKron}. Adapting the notation from \cite{EstCovMatKron}, define the mappings $\hat{\mathbf{A}}(\cdot), \hat{\mathbf{B}}(\cdot)$:
\begin{align}
	\underbrace{\hat{\mathbf{A}}(\mathbf{B})}_{p\times p} &= \frac{1}{f} \sum_{k,l=1}^f{[\mathbf{B}^{-1}]_{k,l}\overline{\hat{\mathbf{S}}_n}(l,k)}, \label{A_update} \\
	\underbrace{\hat{\mathbf{B}}(\mathbf{A})}_{f\times f} &= \frac{1}{p} \sum_{i,j=1}^p{[\mathbf{A}^{-1}]_{i,j} \hat{\mathbf{S}}_n(j,i)}, \label{B_update}
\end{align}
where $\overline{\hat{\bS}}_n=\bK_{p,f}^T \hat{\bS}_n \bK_{p,f}$ (see Sec. \ref{sec: notation} for definition of $K_{p,f}$). For fixed $\bB\in S_{++}^f$, $\hat{\bA}(\bB)$ in (\ref{A_update}) is the minimizer of $J(\bA^{-1},\bB^{-1})$ over $\bA\in S_{++}^p$. A similar interpretation holds for (\ref{B_update}). The flip-flop algorithm starts with some arbitrary p.d. matrix $\bA_{init}$ and computes $\bB$ using (\ref{B_update}), then $\bA$ using (\ref{A_update}), and repeats until convergence. This algorithm does not account for sparsity. 

If $\bTheta_0=\bX_0\otimes \bY_0$ is a sparse matrix, which implies that at least one of $\bX_0$ or $\bY_0$ is sparse, one can penalize the outputs of the flip-flop algorithm and minimize 
\begin{equation} \label{J_lambda_func}
	J_\lambda(\bX,\bY) = J(\bX,\bY) + \bar{\lambda}_X |\bX|_1 + \bar{\lambda}_Y |\bY|_1 .
\end{equation}
This leads to an algorithm that we call KGlasso (see Algorithm \ref{alg: algKGL}), which sparsifies the Kronecker factors in proportion to the parameters $\bar{\lambda}_X, \bar{\lambda}_Y >0$.

\begin{algorithm}
\caption{Kronecker Graphical Lasso (KGlasso) }
\label{alg: algKGL}
\begin{algorithmic}[1]
\STATE \textbf{Input:}  {$\hat{\bS}_n$, $p$, $f$, $n$, $\bar{\lambda}_X >0$, $\bar{\lambda}_Y >0$}
\STATE \textbf{Output:} {$\hat{\bTheta}_{KGlasso}$}

	\STATE Initialize $\bA_{init}$ to be positive definite satisfying Assumption \ref{assumption_posdef_unif}.
	\STATE $\check{\bX} \leftarrow \bA_{init}^{-1}$

	\REPEAT 
	  \STATE	$\hat{\bB} \leftarrow \frac{1}{p} \sum_{i,j=1}^p{[\check{\bX}]_{i,j} \hat{\bS}_n(j,i)}$ (see Eq. (\ref{A_update}))
  	\STATE  $\check{\bY} \leftarrow \bG(\hat{\bB}, \frac{\bar{\lambda}_Y}{p})$, where $\bG(\cdot,\cdot)$ is defined in (\ref{G_operator})
  	\STATE  $\hat{\bA} \leftarrow \frac{1}{f} \sum_{k,l=1}^f{[\check{\bY}]_{k,l} \overline{\hat{\bS}_n}(l,k)}$ (see Eq. (\ref{B_update}))
  	\STATE  $\check{\bX} \leftarrow \bG(\hat{\bA}, \frac{\bar{\lambda}_X}{f})$
	\UNTIL {convergence}
	
	\STATE $\hat{\bTheta}_{KGlasso} \leftarrow \check{\bX} \otimes \check{\bY}$
\end{algorithmic}
\end{algorithm}

The Glasso mapping (\ref{opt_prob}) is written as $\bG(\cdot,\lambda): S^d \to S^d$,
\begin{equation} \label{G_operator}
	\bG(\bT,\lambda) = \arg\min_{\bTheta \in S_{++}^d} \Big\{ \tr(\bTheta \bT)-\log\det(\bTheta)+\lambda|\bTheta|_1 \Big\}.
\end{equation}
As compared to the $\mathcal O(p^4f^4)$ computational complexity of Glasso, KGlasso has a computational complexity of only $\mathcal{O}(p^4+f^4)$ \footnote{In the sparse Kronecker factor case, this cost can be reduced to $\mathcal{O}(p^3+f^3)$.}. 

\section{Convergence of KGlasso Iterations} \label{sec: convergence_analysis_KGL}
In this section, we provide an alternative characterization of the KGlasso algorithm and prove convergence to a local minimum of the objective function.

\subsection{Block-Coordinate Reformulation of KGlasso}

The KGlasso algorithm can be re-formulated as a block-coordinate optimization of the penalized objective function \ref{J_lambda_func}. 

\begin{lemma} \label{dual_lemma}
\begin{enumerate}
	\item Assume $\lambda_X,\lambda_Y \geq 0$ and $\bX\in S_{++}^p,\bY\in S_{++}^f$. When one argument of $J_\lambda(\bX,\bY)$ is fixed, the objective function (\ref{J_lambda_func}) is convex in the other argument.
	\item Assume $\hat{\bS}_n$ is positive definite. Consider $J_\lambda(\mathbf{X},\mathbf{Y})$ in (\ref{J_lambda_func}) with matrix $\mathbf{X}\in S_{++}^p$ fixed. Then, the dual subproblem for minimizing $J_\lambda(\mathbf{X},\mathbf{Y})$ over $\mathbf{Y}$ is:
\begin{equation} \label{dualY}
	\max_{ |\mathbf{W}-\frac{1}{p}\sum_{i,j=1}^p{\mathbf{X}_{i,j} \hat{\bS}_n(j,i)}|_{\infty} \leq \lambda_Y }{ \log \det(\mathbf{W})  }
\end{equation}
where $\lambda_Y := \bar{\lambda}_Y/p$.

On the other hand, consider (\ref{J_lambda_func}) with matrix $\mathbf{Y} \in S_{++}^f$ fixed. Then, the dual problem for minimizing $J_\lambda(\mathbf{X},\mathbf{Y})$ over $\bX$ is:
\begin{equation} \label{dualX}
	\max_{ |\mathbf{Z}-\frac{1}{f}\sum_{k,l=1}^f{\mathbf{Y}_{k,l} \overline{\hat{\bS}_n}(l,k)}|_{\infty}\leq \lambda_X  }{ \log \det(\mathbf{Z})  }
\end{equation}
where $\overline{\hat{\bS}_n}:=\bK_{p,f}^T \hat{\bS}_n \bK_{p,f}$ and $\lambda_X := \bar{\lambda}_X/f$.
	
	\item Strong duality holds for (\ref{dualY}) and (\ref{dualX}).
	\item The solutions to (\ref{dualY}) and (\ref{dualX}) are positive definite.
\end{enumerate}
\end{lemma}
\begin{IEEEproof}
	See Appendix.
\end{IEEEproof}

Note that both dual subproblems (\ref{dualY}) and (\ref{dualX}) have a unique solution and the maximum is attained in each one. This follows from the fact that in each case we are maximizing a strictly concave function over a closed convex set. Lemma \ref{dual_lemma} is similar to the result obtained in \cite{ModelSel}, but with $(\frac{1}{p}\sum_{i,j=1}^p{\bX_{i,j} \hat{\bS}_n(j,i)}, \lambda_Y)$ playing the role of $(\hat{\bS}_n,\lambda)$, for the ``fixed $\bX$'' subproblem.

\subsection{Limit Point Characterization of KGlasso}

We will first show that KGlasso converges to a fixed point. Let $J_\lambda(\bX,\bY)$ be as defined in (\ref{J_lambda_func}) and define $J_\lambda^{(k)}=J_\lambda(\mathbf{X}^{(k)},\mathbf{Y}^{(k)})$ for $k=0,1,2,\dots$.

\begin{theorem} \label{convergence_fixed_point}
	If $n \geq \max(\frac{p}{f},\frac{f}{p})+1$, KGlasso converges to a fixed point. Also, we have $J_\lambda^{(k)} \searrow J_\lambda^{(\infty)}$.
\end{theorem}
\begin{IEEEproof}
	See Appendix.
\end{IEEEproof}

The following analysis uses Theorem \ref{convergence_fixed_point} to prove convergence of the KGlasso algorithm to a local minimum. To do this, we consider a more general setting. The KGlasso algorithm is a special case of Algorithm \ref{alg: alg2}. Assuming a $k$-fold Kronecker product structure for the covariance matrix, the optimization problem (\ref{J_lambda_func}) can be written in the form:
\begin{equation} \label{obj_func_matrix}
	J_\lambda(\bX_1,\dots,\bX_k) = J_0(\bX_1,\dots,\bX_k) + \sum_{i=1}^k{J_i(\bX_i) + \bar{\lambda}_i \eta_1(\bX_i)}
\end{equation}
where $\bX_i \in S_{++}^{d_i}$, $\eta_1(\bX_i):=|\bX_m|_1$, $J_0(\bX_1,\dots,\bX_k):=\tr((\bX_1\otimes \bX_2 \otimes \dots \otimes \bX_k)\hat{\bS}_n)$ and $J_i(\bX_i) = - \prod_{i' \neq i}{d_{i'}} \cdot \log\det(\bX_i)$ for $i=1,\dots,k$.

Without loss of generality, by reshaping matrices into appropriate vectors, (\ref{obj_func_matrix}) can be rewritten as:
\begin{equation} \label{general_f}
	J_\lambda(\mathbf{x}_1,\dots,\mathbf{x}_k) = J_0(\mathbf{x}_1,\dots,\mathbf{x}_k) + \sum_{i=1}^k{J_i(\mathbf{x}_i)+\bar{\lambda}_i \eta_i(\mathbf{x}_i)}
\end{equation}
where the optimization variable is $\mathbf{x}:=[\mathbf{x}_1^T, \mathbf{x}_2^T,\dots,\mathbf{x}_k^T]^T \in \RR^{d'}$, where $\mathbf{x}_i\in \RR^{d_i^2}$ and $d'=\sum_{i=1}^k{d_i^2}$. For example, $\eta_i(\mathbf{X}_i)=|\mathbf{X}_i|_1=\nn \vec(\mathbf{X}_i) \nn_1 = \nn \mathbf{x}_i\nn_1 = \eta_i(\mathbf{x}_i)$. The mapping $\{J_i\}_{i=0}^k$ can be similarly written in terms of the vectors $\mathbf{x}_i$ instead of the matrices $\mathbf{X}_i$.

The reader can verify that the objective function (\ref{obj_func_matrix}) satisfies the properties (for $n\geq \max(\frac{p}{f},\frac{f}{p}) + 1$) in Appendix D.

The general optimization problem of interest here is:
\begin{equation} \label{gen_func}
	\min_{\mathbf{x} \in \RR^{d'}} {J_\lambda(\mathbf{x})} \text{ subject to } \vec^{-1}(\mathbf{x}_i)=\bX_i \in S_{++}^{d_i}, i=1,\dots,k
\end{equation}
The positive definiteness constraints are automatically taken care of by the construction of the algorithm (see Lemma \ref{dual_lemma}.4). Let the dimension of the covariance matrix be denoted by $d:=\prod_{i=1}^k{d_i}$. We assume $n>d$. To solve (\ref{gen_func}), a block coordinate-descent penalized algorithm is constructed:

\begin{algorithm}
\caption{Block Coordinate-Descent Penalized Algorithm}
\label{alg: alg2}
\begin{algorithmic}[1]

\STATE \textbf{Input:}  {$\hat{\bS}_n$, $d_i$, $n$, $\epsilon>0$, $\lambda_i >0$}
\STATE \textbf{Output:} {$\hat{\bTheta}$}

	\STATE Initialize $\mathbf{X}_1^0, \mathbf{X}_2^0, \dots, \mathbf{X}_k^0$ matrices as positive definite matrices, e.g., scaled identity.
	
	\STATE $\hat{\mathbf{\Theta}}_0 \leftarrow \mathbf{X}_1^0\otimes \mathbf{X}_2^0 \otimes \dots \otimes \mathbf{X}_k^0$\;
	
	\STATE $m \leftarrow 0$\;
	\REPEAT {
		\STATE $\hat{\mathbf{\Theta}}_{\text{prev}} \leftarrow \hat{\mathbf{\Theta}}$
		
		\STATE $\mathbf{X}_1^m \leftarrow \arg \min_{\mathbf{A}_1 \succ 0} J_\lambda(\mathbf{A}_1,\mathbf{X}_2^{m-1},\dots,\mathbf{X}_k^{m-1})$
		\STATE $\mathbf{X}_2^m \leftarrow \arg \min_{\mathbf{A}_2 \succ 0} J_\lambda(\mathbf{X}_1^m,\mathbf{A}_2,\dots,\mathbf{X}_k^{m-1})$
		\STATE $\vdots$
		\STATE $\mathbf{X}_k^m \leftarrow \arg \min_{\mathbf{A}_k \succ 0} J_\lambda(\mathbf{X}_1^m,\mathbf{X}_2^m,\dots,\mathbf{A}_k)$
		
		\STATE $\hat{\mathbf{\Theta}} \leftarrow \mathbf{X}_1^m \otimes \mathbf{X}_2^m \otimes \dots \otimes \mathbf{X}_k^m$
		
		\STATE $m \leftarrow m + 1$
	}	\UNTIL {$\nn \hat{\mathbf{\Theta}}_{\text{prev}}-\hat{\mathbf{\Theta}} \nn \leq \epsilon$}
	
\end{algorithmic}
\end{algorithm}

\begin{remark}
	The positive definiteness constraint at each coordinate descent iteration of Algorithms \ref{alg: algKGL} and \ref{alg: alg2} need not be explicit since the objective function $J_\lambda(\cdot)$ acts as a logarithmic barrier function.
\end{remark}

Note that Algorithm \ref{alg: algKGL} is a special case of Algorithm \ref{alg: alg2}. An extension of Theorem \ref{convergence_fixed_point}, assuming $n>d$ or $J_\lambda^*> -\infty$, based on induction, can be used to show that the limit points of the sequence of iterates $(\mathbf{x}^m)_{m\geq 0}=(\mathbf{x}_1^m,\dots,\mathbf{x}_k^m)_{m \geq 0}$ are fixed points.

\begin{remark}
	Note that a necessary condition for $\mathbf{x}^*$ to minimize $J_\lambda$ is $0\in \partial J_\lambda(\mathbf{x}^*)$. This is not sufficient however.
\end{remark}

We next show that the limit point(s) of $(\mathbf{x}^m)_{m\geq 0}$ are nonempty and are local minima.

\begin{theorem} \label{convergence_theorem_critical}
	Let $(\mathbf{x}^m)=(\mathbf{x}_1^m,\dots,\mathbf{x}_k^m)_{m\geq 0}$ be a sequence generated by Algorithm \ref{alg: alg2}. Assume $n>d$ \footnote{This requirement on the sample size can be significantly relaxed. For the two-fold case, this can be relaxed to $n\geq \max(\frac{p}{f},\frac{f}{p})+1$.}.
	
\begin{enumerate}
	\item The algorithm converges to a local minimum.
	\item If $\mathbf{x}^0$ is not a local minimum, strict descent follows.
\end{enumerate}
\end{theorem}
\begin{IEEEproof}
	See Appendix.
\end{IEEEproof}

As a consequence of Theorem \ref{convergence_theorem_critical}, we have the following corollary.

\begin{corollary} \label{cor_localmin_singleton}
	Assuming $n\geq \max(\frac{p}{f},\frac{f}{p})+1$, the KGlasso algorithm converges to a local minimizer of the objective function (\ref{J_lambda_func}).
\end{corollary}

\section{High Dimensional Consistency of FF} \label{sec: FF_rate}
In this section, we show that the flip-flop (FF) algorithm achieves the optimal (non-sparse) statistical convergence rate of $O_P \left(\sqrt{\frac{p^2+f^2}{n}}\right)$. This result (see Thm. \ref{thm: FF_optimal_rate}) allows us to establish that the proposed KGlasso has significantly improved MSE convergence rate (see Thm. \ref{thm: KGL_optimal_rate}). We make the following standard assumption on the spectra of the Kronecker factors.

\begin{assumption} \label{assumption_posdef_unif}
Uniformly Bounded Spectra \\
There exist absolute constants $\underline{k}_A, \overline{k}_A, \underline{k}_B, \overline{k}_B, \underline{k}_{A_{init}}, \overline{k}_{A_{init}}$ such that: \\
\indent 1a. $0<\underline{k}_A \leq \lambda_{min}(\bA_0) \leq \lambda_{max}(\bA_0) \leq \overline{k}_A < \infty$ \\
\indent 1b. $0<\underline{k}_B \leq \lambda_{min}(\bB_0) \leq \lambda_{max}(\bB_0) \leq \overline{k}_B < \infty$ \\
\indent 2. $0<\underline{k}_{A_{init}} \leq \lambda_{min}(\bA_{init}) \leq \lambda_{max}(\bA_{init}) \leq \overline{k}_{A_{init}} < \infty$
\end{assumption}

Let $\bSigma_{FF}(3):=\hat{\bA}(\hat{\bB}(\bA_{init})) \otimes \hat{\bB}(\hat{\bA}(\hat{\bB}(\bA_{init})))$ denote the 3-step (noniterative) version of the flip-flop algorithm \cite{EstCovMatKron}. More generally, let $\bSigma_{FF}(k)$ denote the $k$-step version of the flip-flop algorithm, and denote its inverse as $\bTheta_{FF}(k)=(\bSigma_{FF}(k))^{-1}$.

%

\begin{theorem} \label{thm: FF_optimal_rate}
	Let $\bA_0,\bB_0$, and $\bA_{init}$ satisfy Assumption \ref{assumption_posdef_unif} and define $M=\max(p,f,n)$. Assume $p\geq f\geq 2$ and $p \log M \leq C'' n$ for some finite constant $C''>0$. Finally, assume $n \geq \frac{p}{f} + 1$.
	Then, for $k\geq 2$ finite,
	\begin{equation} \label{FF_rate_2}
		\nn \bTheta_{FF}(k) - \bTheta_0 \nn_F = O_P\left( \sqrt{\frac{(p^2+f^2) \log M}{n}} \right)
	\end{equation}
	as $n\to\infty$.
\end{theorem}
\begin{IEEEproof}
	See Appendix.
\end{IEEEproof}
\begin{remark}
	The sufficient conditions are symmetric with respect to $p$ and $f$-i.e. for $f\geq p$, the corresponding conditions would become $f\log M\leq C'' n$ for some constant $C''>0$, and $n \geq \frac{f}{p} + 1$.
\end{remark}


To achieve accurate covariance estimation for arbitrarily structured Kronecker factors, the minimal sample size needed is $n=\Omega((p^2+f^2)\log M)$.

The bound (\ref{FF_rate_2}) specifies the rate of reduction of the estimation error for the multi-iteration FF algorithm, which includes the three step FF algorithm ($k=3$) \cite{EstCovMatKron} as a special case. The error reduction decreases as long as $p$ and $f$ do not increase too quickly in $n$.

Note that (\ref{FF_rate_2}) specifies a faster rate than that of the naive sample covariance matrix estimator (\ref{naive_SCM_rate}). Furthemore, since the computational complexity for FF is $\mathcal{O}(p^2+f^2)$ which is less than the $\mathcal{O}(p^2f^2)$ complexity of SCM, by exploiting Kronecker structure FF simultaneously achieves improved MSE performance and reduced computational complexity.

\begin{figure}[ht]
	\centering
		\includegraphics[width=0.50\textwidth]{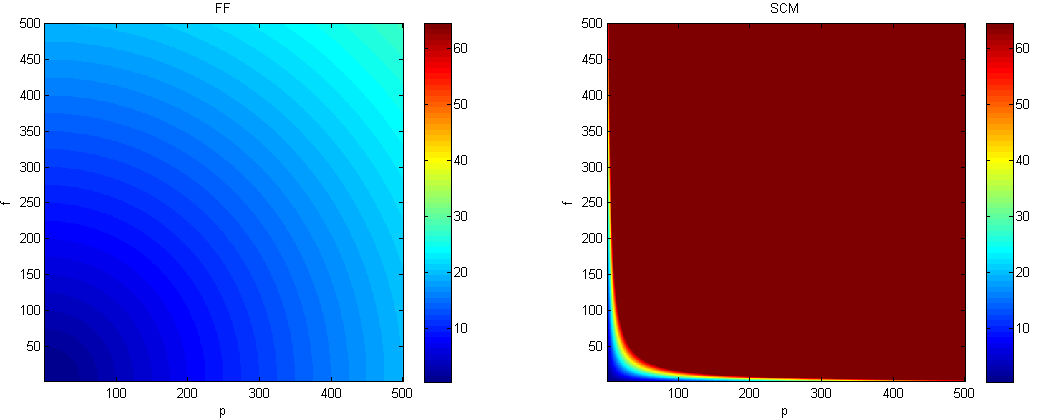}
	\caption{Root mean square error (RMSE) performance for the flip-flop estimator (FF) (left) and for the standard sample covariance matrix estimator (SCM) (right). SCM performs very poorly in comparison to FF when the covariance matrix decomposes as a Kronecker product. Here, the sample size $n$ is fixed and the dimensions of the Kronecker factors ($p,f$) vary. Equation (\ref{FF_rate_2}) is plotted on the left and Equation (\ref{naive_SCM_rate}) on the right. Exploiting structure yields a significant reduction in MSE. The magnitude of the colormap reflects the error up to a constant scaling. The colormap in both images is the same, which visually shows the lower RMSE of FF as compared to SCM. }
	\label{fig: FF_vs_SCM}
\end{figure}

\section{High Dimensional Consistency of KGlasso} \label{sec: consistency}
In this section, consistency is established for KGlasso as $p,f,n\to\infty$.

\subsection{MSE convergence rate of KGlasso}
Define $\bTheta_{KGlasso}(k)$ as the output of the $k$th compression and sparsification step (two of these steps constitute a full KGlasso iteration).

\begin{theorem} \label{thm: KGL_optimal_rate}
	Let $\bA_0,\bB_0, \bA_{init}$ satisfy Assumption \ref{assumption_posdef_unif}. Let $M=\max(p,f,n)$.	Let $\bar{\lambda}_Y^{(1)} \asymp p \sqrt{\frac{\log M}{np}}$ and $\bar{\lambda}_X^{(k)} \asymp \left( \frac{1}{\sqrt{p}} + \frac{1}{\sqrt{f}} \right) f \sqrt{\frac{\log M}{n}}, \bar{\lambda}_Y^{(k')} \asymp \left( \frac{1}{\sqrt{p}} + \frac{1}{\sqrt{f}} \right) p \sqrt{\frac{\log M}{n}}$ as $p,f,n\to\infty$ for all $k \geq 1$ and $k' \geq 2$. Assume sparse $\bX_0$ and $\bY_0$, i.e. $s_{X_0}=O(p), s_{Y_0}=O(f)$. Assume $\max\left( \frac{p}{f}, \frac{f}{p}\right) \log M=o(n)$. Then, for $k\geq 2$ finite, we have
	\begin{equation} \label{KGL_perfect_rate}
		\nn \bTheta_{KGlasso}(k) - \bTheta_0 \nn_F = O_P\left( \sqrt{\frac{(p+f) \log M}{n}} \right)
	\end{equation}
	as $p,f,n\to\infty$.
\end{theorem}
\begin{IEEEproof}
	See Appendix.
\end{IEEEproof}

Theorem \ref{thm: KGL_optimal_rate} offers a strict improvement over standard Glasso \cite{Rothman, ModelSel} and generalizes Thm. 1 in \cite{Rothman} to the case of sparse Kronecker product structure. Thm. \ref{thm: KGL_optimal_rate} generalizes Thm. \ref{thm: FF_optimal_rate} to the case of sparse Kronecker structure. 
Comparison between the error expressions (\ref{Glasso_rate}), (\ref{FF_rate_2}) and (\ref{KGL_perfect_rate}) show that, by exploiting both Kronecker structure and sparsity, KGlasso can attain significantly lower estimation error than standard  Glasso \cite{Rothman} and FF \cite{EstCovMatKron}. To achieve accurate covariance estimation for the sparse Kronecker product model, the minimal sample size needed is $n=\Omega((p+f)\log M)$.

Although Thm. \ref{thm: KGL_optimal_rate} shows a rate on the inverse covariance matrix, this asymptotic rate can be shown to hold for the covariance matrix as well (i.e., the inverse of $\bTheta_{KGlasso}$). 

\begin{figure}[ht]
	\centering
		\includegraphics[width=0.50\textwidth]{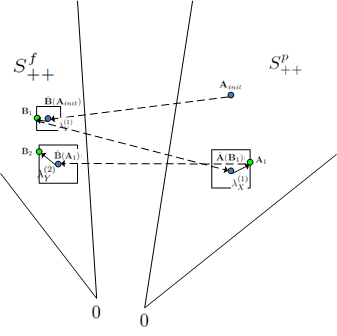}
	\caption{ Illustration of first three iterations of KGlasso. The squares around the blue dots represent the $\ell_\infty$ balls controlled by the regularization parameter (see dual programs (\ref{dualY}) and (\ref{dualX})). As the regularization parameters tend to zero, the balls shrink to the blue points, and KGlasso becomes identical to the FF algorithm. }
	\label{fig: KGL_visual}
\end{figure}

Let $\bB_1:=\bG(\hat{\bB}(\bA_{init}),\lambda_Y^{(1)})^{-1}$, where $\bG$ is defined in (\ref{G_operator}). Then, $\bTheta_{KGlasso}(1) = \bG(\hat{\bA}(\bB_1),\lambda_X^{(1)}) \otimes \bG(\hat{\bB}(\bA_{init}),\lambda_Y^{(1)})$ denotes the KGlasso output after the the first two steps of the KGlasso algorithm (or one KGlasso iteration). A graphical depiction of the first three steps of KGlasso is shown in Fig. \ref{fig: KGL_visual}. Define $\bB_1=\bG(\hat{\bB}(\bA_{init}),\lambda_Y^{(1)})^{-1}$, where $\bG$ is given in (\ref{G_operator}). Then, $\bTheta_{KGlasso}(1) = \bG(\hat{\bA}(\bB_1),\lambda_X^{(1)}) \otimes \bG(\hat{\bB}(\bA_{init}),\lambda_Y^{(1)})$ denotes the KGlasso output after the the first two steps of the KGlasso algorithm (or one KGlasso iteration). Although Thm. \ref{thm: KGL_optimal_rate} shows a rate on the inverse covariance matrix, this asymptotic rate can be shown to hold for the covariance matrix as well (see proof of Thm. \ref{thm: KGL_optimal_rate} in Appendix).

\begin{figure}[ht]
	\centering
		\includegraphics[width=0.50\textwidth]{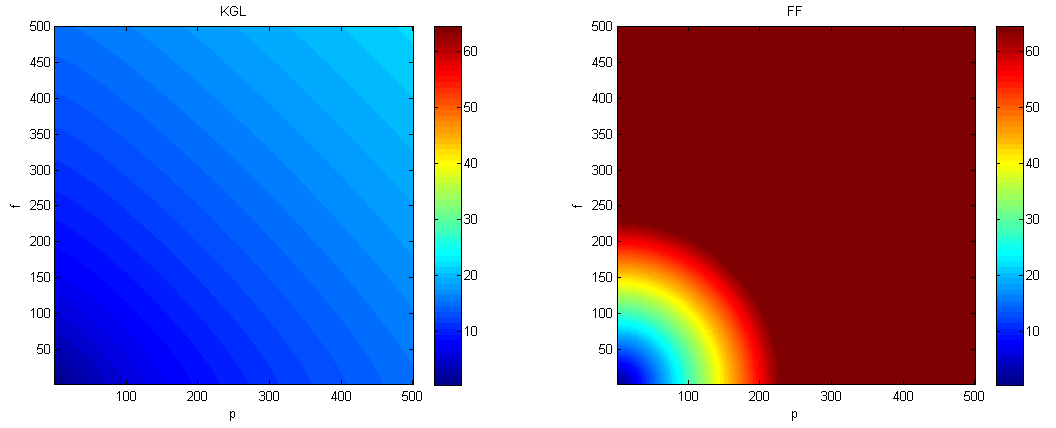}
	\caption{Root mean square error performance for Kronecker graphical lasso estimator (KGlasso) (left) and flip-flop estimator (FF) (right). FF performs very poorly in comparison to KGlasso when the covariance matrix decomposes as a Kronecker product and both Kronecker factors are sparse. The bound in Equation (\ref{KGL_perfect_rate}) is plotted on the left and that in Equation (\ref{FF_rate_2}) on the right. The magnitude of the colormap reflects the error up to a constant scaling. }
	\label{fig: KGL_vs_FF}
	\vfill
		\includegraphics[width=0.50\textwidth]{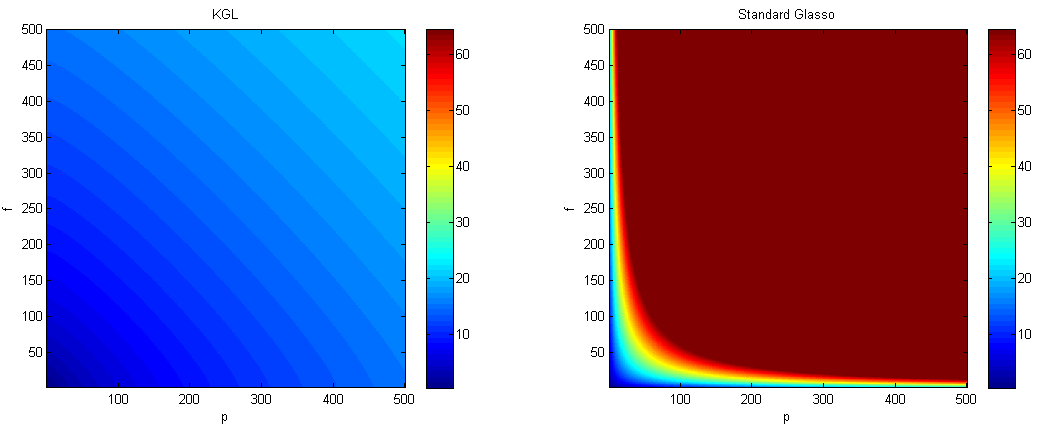}
	\caption{Root mean square error performance for Kronecker graphical lasso estimator (KGlasso) (left) and standard Glasso estimator (Glasso) (right). Glasso performs very poorly in comparison to KGlasso when the covariance matrix decomposes as a Kronecker product and both Kronecker factors are sparse. The bound in Equation (\ref{KGL_perfect_rate}) is plotted on the left and that in Equation (\ref{Glasso_rate}) on the right. The magnitude of the colormap reflects the error up to a constant scaling. }
	\label{fig: KGL_vs_Glasso}
\end{figure}

Figures \ref{fig: KGL_vs_FF} and \ref{fig: KGL_vs_Glasso} graphically compare the MSE convergence rates of KGlasso, FF and standard Glasso as a function of $p,f$ for fixed $n$. Note that the standard Glasso algorithm would yield an inferior rate to (\ref{KGL_perfect_rate}) (recall (\ref{Glasso_rate})).

\begin{figure}[ht]
	\centering
		\includegraphics[width=0.50\textwidth]{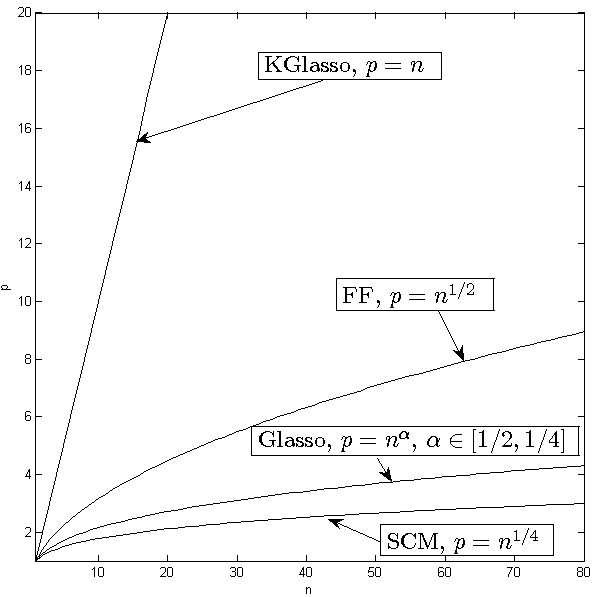}
	\caption{ Graphical depiction of minimal sample size required for KGlasso, FF, Glasso and naive SCM estimators to achieve accurate covariance estimation. The region below the lines constitute the MSE convergence regions-i.e. traveling along a path $(p(n),n)$ as $n\to\infty$ within such regions implies the MSE convergence rate tends to zero (see (\ref{naive_SCM_rate}),(\ref{Glasso_rate}),(\ref{FF_rate_2}) and (\ref{KGL_perfect_rate})). }
	\label{fig: visRate}
\end{figure}

The minimal sample size required to achieve accurate covariance estimation is graphically depicted in Fig. \ref{fig: visRate} for the special case $p=f$. The regions below the lines are the MSE convergence regions-i.e., the MSE convergence rate goes to zero as $p,n$ grow together to infinity at a certain growth rate controlled by these regions. It is shown that KGlasso allows the dimension $p$ to grow almost linearly in $n$ and still achieve accurate covariance estimation (see (\ref{KGL_perfect_rate})) and thus, uniformly outperforms FF, Glasso and the naive SCM estimators in the case both Kronecker factors are sparse.

\subsection{Discussion}
Theorem \ref{thm: KGL_optimal_rate} is established using the large deviation bound in Lemma \ref{lemma: large_dev_optimal}. We provide some intuition on this bound below. 
Assume that $\bX_{init} = \bX_0$, or $\bA_{init}=\bX_{init}^{-1}=\bA_0$. Define $\bW= \bX_0^{1/2} \otimes \bI_p$ and $\tilde{\bz}_t = \bW \bz_t$, with i.i.d. $\bz_t\sim N(\mathbf{0}, \bA_0\otimes \bB_0)$, $t=1,\dots,n$. Then, $\tilde{\bz}_t$ has block-diagonal covariance
\begin{equation*}
	\Cov(\tilde{\bz}_t) = \bI_p \otimes \bB_0.
\end{equation*}
When $\bW$ is applied to the transformed $pf\times pf$ sample covariance matrix, $\hat{\bS}_n^W := \bW \hat{\bS}_n \bW^T$, the first step of KGlasso produces an iterate $\hat{\bY}_n^{(1)}=\bG(\hat{\bB},\lambda_Y)$ with $\hat{\bB} = \frac{1}{p}\sum_{i=1}^p \hat{\bS}_n^W(i,i)$ (recall (\ref{B_update})). For suitable $\lambda_Y = \lambda_Y^{(1)}$, $\hat{\bY}_n^{(1)}$ converges to $\bY_0$ with respect to maximal elementwise norm at a rate $O_P\left( \sqrt{\frac{\log M}{np}} \right)$. The convergence of $\hat{\bY}_n^{(1)}$ is easily established by applying the Chernoff bound and invoking the jointly Gaussian property of the measurements and the block diagonal structure of $\Cov(\tilde{\bz}_t)$. Lemma \ref{lemma: large_dev_optimal} in the Appendix establishes that this rate holds even if $\bX_{init}\neq \bX_0$ in Assumption \ref{assumption_posdef_unif}. In view of the rate of convergence of $\hat{\bY}^{(1)}$, to achieve a reduction in the MSE of $\bY$, either the sample size $n$ or the dimension $p$ must increase. Lemma \ref{lemma: large_dev_optimal} provides a tight bound that makes the dependence of the convergence rate explicit in $p,f$ and $n$. Theorem \ref{thm: KGL_optimal_rate} uses Lemma \ref{lemma: large_dev_optimal} to show that KGlasso converges to $\bX_0\otimes \bY_0$ with rate $O_P\left( \sqrt{\frac{(p+f)\log M}{n}} \right)$ with respect to Frobenius norm.

\section{Simulation Results} \label{sec: simulations}

In this section, we empirically validate the convergence rates established in previous sections using Monte Carlo simulation. 

Each iteration of the KGlasso involves solving an $\ell_1$ penalized covariance estimation problem of dimension $100 \times 100$ (Step 6 and Step 8 of KGlasso specified by Algorithm \ref{alg: algKGL}). To solve these small sparse covariance estimation problems we used the Glasso algorithm of Hsieh {\it et al} \cite{HsiehNIPS11} where the Glasso stopping criterion was determined by monitoring when the duality gap falls below a threshold of $10^{-3}$.

To evaluate performance, Monte Carlo simulations were used. Unless otherwise specified, the true matrices $\bX_0:=\bA_0^{-1}$ and $\bY_0:=\bB_0^{-1}$ were unstructured randomly generated positive definite matrices based on an Erd\"{o}s-R\'{e}nyi graph model. First, a square binary matrix $\bC$ was generated based on independently and identically distributing ``0s'' with a probability $p^*$ and ``1s'' with a probability $1-p^*$. Then, $\tilde{\bC} := (\bC + \bC^T)/2$ symmetrizes the matrix. The perturbation level $\rho$ was selected as $\rho = 0.05-\lambda_{min}(\tilde{\bC})$, producing $\bY_0 := \tilde{\bC} + \rho \bI_f$, the sparse inverse matrix. There was a total of $20$ trial runs for each fixed number of samples $n$. Performance assessment was based on normalized Frobenius norm error in the covariance and precision matrix estimates. The normalized error was calculated using
\begin{equation*}
	\sqrt{ \frac{1}{N_{MC}} \sum_{i=1}^{N_{MC}}{ \frac{\nn \bSigma_{0}-\hat{\bSigma}(i) \nn_F^2}{\nn \bSigma_0 \nn_F^2} } }
\end{equation*}
where $N_{MC}$ is the number of Monte Carlo runs and $\hat{\bSigma}(i)$ is the covariance output from the $i$th trial run. The same formula can be adapted to calculate the normalized error in the precision matrix $\hat{\bTheta}_0$.
In the implementation of KGlasso, the regularization parameters were chosen as follows. The initialization was $\bX_{init}=\bI_p$. The regularization parameters were selected as $\lambda_Y^{(1)}= c_y \sqrt{\frac{\log M}{np}}$, $\lambda_X^{(2)}= c_x \sqrt{\frac{\log M}{nf}} + \lambda_Y^{(1)}$, $\lambda_Y^{(2)} = \lambda_X^{(2)}$, $\lambda_X^{(3)}= \lambda_X^{(2)}$, etc. For Examples 1 and 2 below, the (positive) scaling constants $(c_x,c_y)$ in front of the regularization parameters were chosen experimentally to optimize respective performances. For Example 3, we simply set $c_x=c_y=0.4$.

\subsection{Example 1} 
\begin{figure}[htp]
	\centering
		\includegraphics[width=0.40\textwidth]{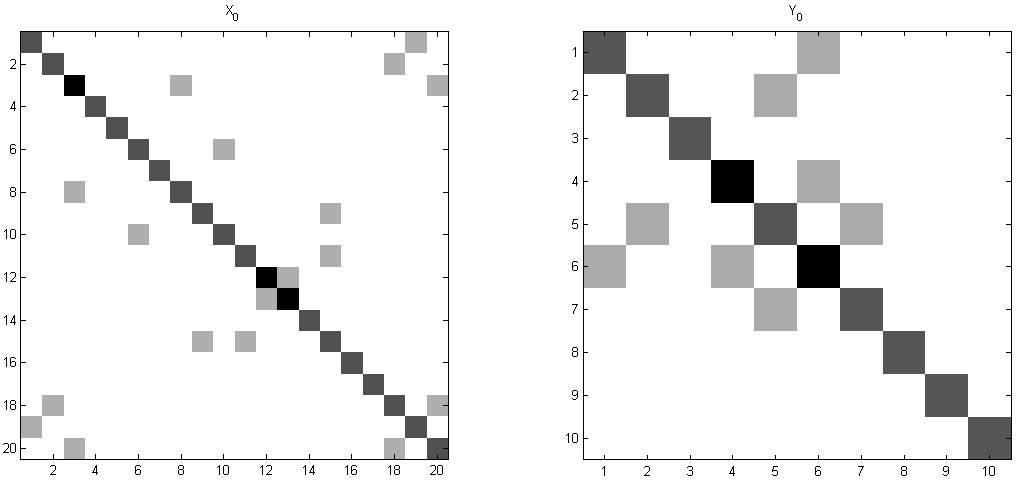}
	\caption{Doubly sparse Kronecker matrix representation for simulation example 1. Left panel: left Kronecker factor. Middle panel: right Kronecker factor. Right panel: Kronecker product inverse covariance matrix}
	\label{fig: sim1_matrices}
	\vfill
		\includegraphics[width=0.40\textwidth]{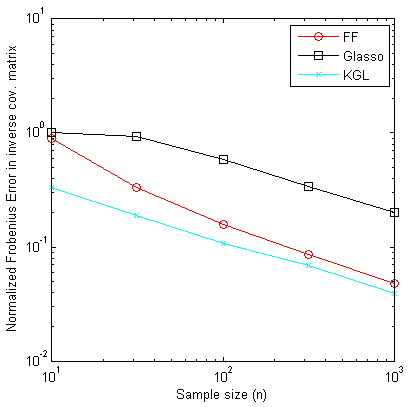}
	\caption{Normalized RMSE of precision matrix estimate $\hat{\bTheta}=\hat{\bSigma}^{-1}$ as a function of sample size $n$ for structure exhibited in Fig. \ref{fig: sim1_matrices}. KGlasso (Kronecker graphical lasso) uniformly outperforms FF (flip-flop) algorithm and standard Glasso algorithm for all $n$. Here, $p=20$ and $f=10$.}
	\label{fig: sim1_frob_inv}
	\vfill
		\includegraphics[width=0.40\textwidth]{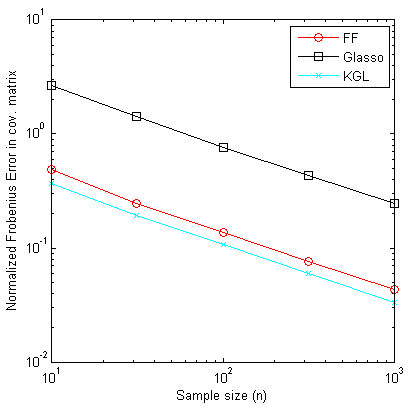}
	\caption{Normalized RMSE of covariance matrix estimate $\hat{\bSigma}$ as a function of sample size $n$ for structure exhibited in Fig. \ref{fig: sim1_matrices}. KGlasso (Kronecker graphical lasso) uniformly outperforms FF (flip-flop) algorithm and standard Glasso algorithm for all $n$. Here, $p=20$ and $f=10$. }
	\label{fig: sim1_frob_cov}
\end{figure}

We consider the simple case that $\bX_0$ and $\bY_0$ are sparse matrices of dimensions $p=20$ and $f=10$. Figure \ref{fig: sim1_matrices} shows that $\bX_0 \otimes \bY_0$ is a perturbation of $\bI_{pf}$. Figures \ref{fig: sim1_frob_inv} and \ref{fig: sim1_frob_cov} compare the root-mean squared error (RMSE) performance in precision and covariance matrices as a function of $n$. As expected, KGlasso outperforms both naive Glasso and FF over the range of $n$ for both the covariance and the inverse covariance estimation problem. As expected, the FF algorithm suffers in the small sample regime. KGlasso outperforms FF in this regime since it exploits sparsity in addition to Kronecker structure.

\subsection{Example 2}
\begin{figure}[htp]
	\centering
		\includegraphics[width=0.40\textwidth]{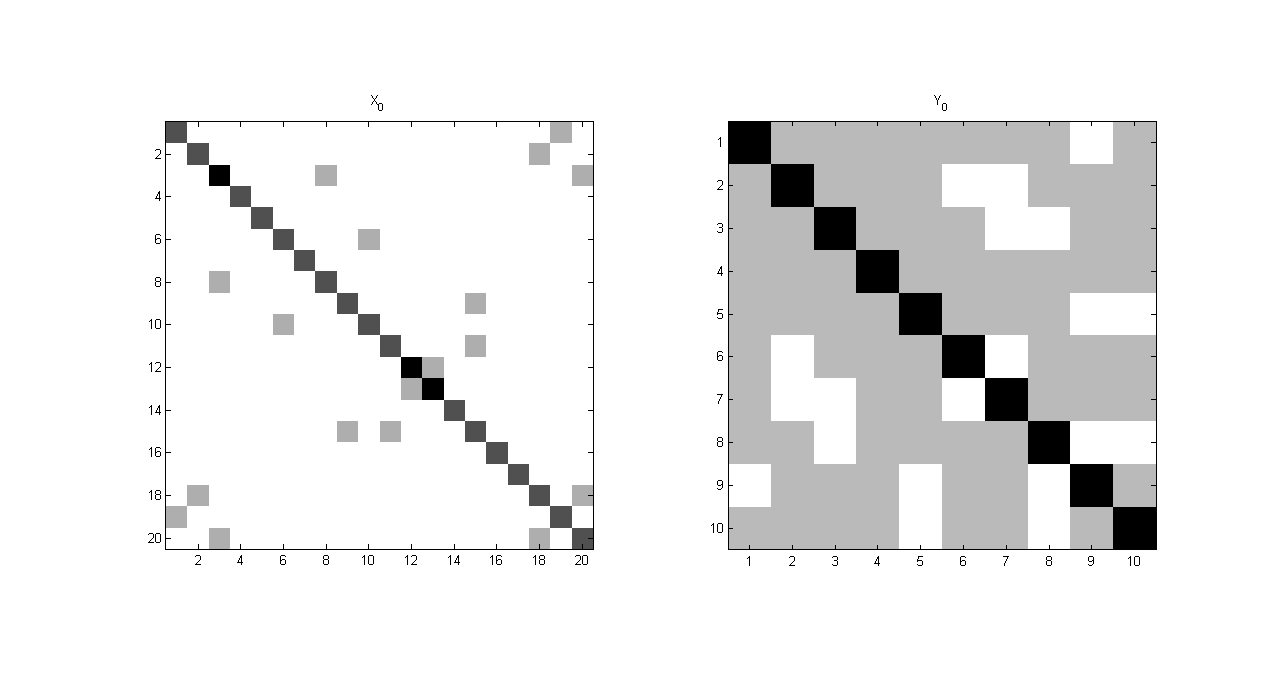}
	\caption{Sparse Kronecker matrix representation for simulation example 2. Left panel: left Kronecker factor. Middle panel: right Kronecker factor. Right panel: Kronecker product inverse covariance matrix.}
	\label{fig: sim2_matrices}
	\vfill
		\includegraphics[width=0.40\textwidth]{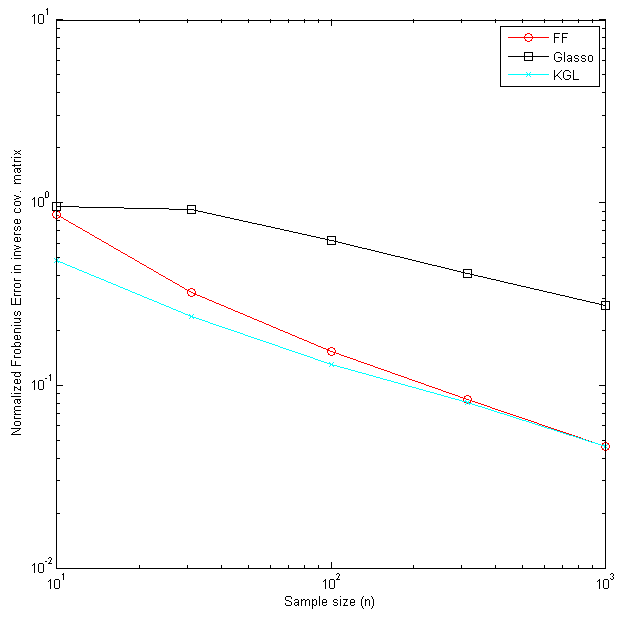}
	\caption{Normalized RMSE performance for precision matrix as a function of sample size $n$. KGlasso (Kronecker graphical lasso) uniformly outperforms FF (flip-flop) algorithm and standard Glasso algorithm for all $n$. Here, $p=20$ and $f=10$. }
	\label{fig: sim2_frob_inv}
	\vfill
		\includegraphics[width=0.40\textwidth]{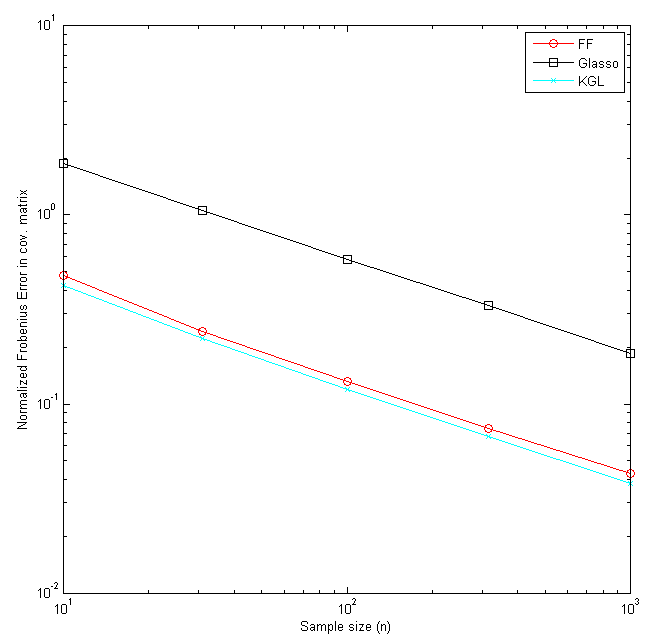}
	\caption{Normalized RMSE performance for covariance matrix as a function of sample size $n$. KGlasso (Kronecker graphical lasso) uniformly outperforms FF (flip-flop) algorithm and standard Glasso algorithm for all $n$. Here, $p=20$ and $f=10$. }
	\label{fig: sim2_frob_cov}
\end{figure}

We consider the case when $\bA_0$ is identity and $\bY_0$ is dense (see Fig. \ref{fig: sim2_matrices}). Figures \ref{fig: sim2_frob_inv} and \ref{fig: sim2_frob_cov} show similar trends to those exhibited in Figures \ref{fig: sim1_frob_inv} and \ref{fig: sim1_frob_cov} for the case that both $\bX_0$ and $\bY_0$ are sparse.

\subsection{Example 3} 
\begin{figure}[htp]
	\centering
		\includegraphics[width=0.40\textwidth]{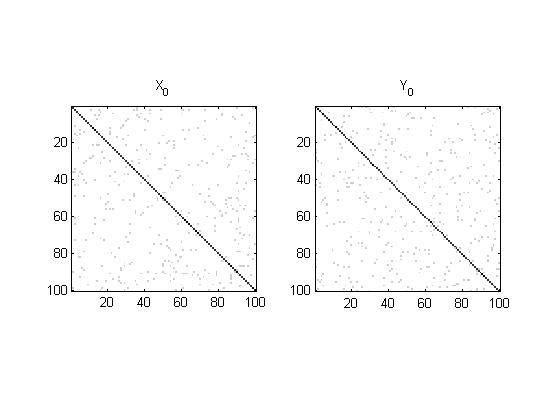}
	\caption{Sparse Kronecker matrix representation. Left panel: left Kronecker factor. Right panel: right Kronecker factor.}
	\label{fig: sim3_matrices}
	\vfill
		 \includegraphics[width=0.40\textwidth]{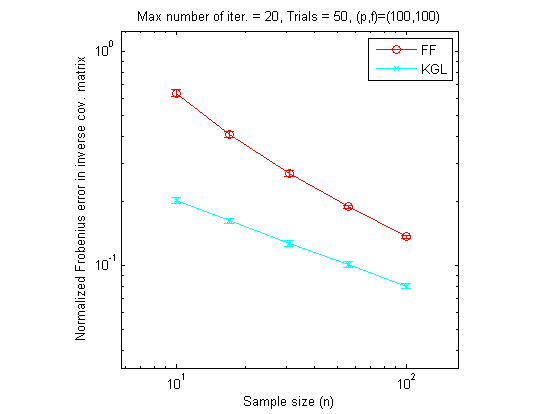}
	\caption{Normalized RMSE performance for precision matrix as a function of sample size $n$. KGlasso (Kronecker graphical lasso) uniformly outperforms FF (flip-flop) algorithm for all $n$. Here, $p=100$ and $f=100$. For $n=10$, there is a $69 \%$ RMSE reduction. }
	\label{fig: sim3_frob_inv}
	\vfill
		 \includegraphics[width=0.40\textwidth]{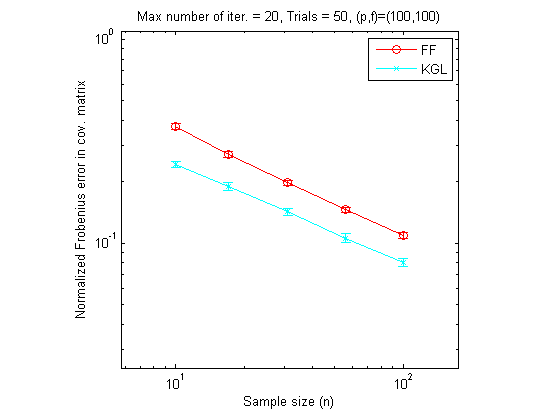}
	\caption{Normalized RMSE performance for covariance matrix as a function of sample size $n$. KGlasso (Kronecker graphical lasso) uniformly outperforms FF (flip-flop) algorithm for all $n$. Here, $p=100$ and $f=100$. For $n=10$, there is a $35 \%$ RMSE reduction. }
	\label{fig: sim3_frob_cov}
\end{figure}

We considered the setting where $\bX_0$ and $\bY_0$ are large sparse matrices of dimension $p=f=100$ (see Fig. \ref{fig: sim3_matrices}). Only 5\% of the off-diagonal entries were nonzero for both matrices $\bX_0$ and $\bY_0$.  The dimension of $\bTheta_0$ is $d=10,000$, which was too large for implementation of standard Glasso. Figures \ref{fig: sim3_frob_inv} and \ref{fig: sim3_frob_cov} compare the root-mean squared
error (RMSE) performance in precision and covariance matrices as a function of $n$. As expected, KGlasso outperforms both naive Glasso and FF over the range of $n$ for both the covariance and the inverse covariance estimation problem. As expected, the FF algorithm suffers in the small sample regime. KGlasso outperforms FF in this regime since it exploits sparsity in addition to Kronecker structure.

For $n=10$, there is a $69 \%$ ($\approx 5.09$ dB) RMSE reduction for the precision matrix and $35 \%$ RMSE reduction for the covariance matrix when using KGlasso instead of FF. For $n=100$, there is a $41 \%$ ($\approx 2.29$ dB) RMSE reduction for the precision matrix and $26 \%$ RMSE reduction for the covariance matrix. For the small sample regime, there is approximately a $5.09$ dB reduction for the precision matrix, which is a significant performance gain.

\subsection{Example 4}
Here, the true covariance matrix factors  $\bX_0=\bA_0^{-1}$ and $\bY_0=\bB_0^{-1}$ were unstructured randomly generated positive definite matrices. First, $p$ random nonzero elements were placed on the diagonal of a square $p \times p$ matrix $C$. Then, on average $p$ nonzero elements were placed on the off-diagonal and symmetry was imposed. On average, a total of $3p$ elements were nonzero. The resulting matrix $\tilde{\bC}$ was regularized to produce the sparse positive definite inverse covariance $\bY_0 = \tilde{\bC} + \rho \bI_f$, where $\rho = 0.5-\lambda_{min}(\tilde{\bC})$.

We also compare KGlasso  to a natural extension of the FF algorithm that accounts  for both sparsity and Kronecker structure. The flip-flop thresholding method (FF/Thres) that we consider consists of first computing the FF solution and then thresholding each estimated precision matrix. To ensure a fair comparison we set the threshold level of FF/Thres that yields exactly the same sparsity factor as the KGLasso estimated precision matrices.

For $n=10$, there is a $72 \%$ ($\approx 5.53$ dB) RMSE reduction for the precision matrix and $41 \%$ RMSE reduction for the covariance matrix when using KGlasso instead of FF. For $n=10$, there is a $70 \%$ ($\approx 5.23$ dB) RMSE reduction for the precision matrix and $62 \%$ RMSE reduction for the covariance matrix when using KGlasso instead of FF/Thres. For $n=100$, there is a $53 \%$ ($\approx 3.28$ dB) RMSE reduction for the precision matrix and $33 \%$ RMSE reduction for the covariance matrix when using KGLasso instead of FF. For $n=100$, there is a $50 \%$ ($\approx 3.01$ dB) RMSE reduction for the precision matrix and $41 \%$ RMSE reduction for the covariance matrix when using KGLasso instead of FF/Thres. For the small sample regime, there is approximately a $5.53$ dB reduction for the precision matrix, which is a significant performance gain.
\begin{figure}[htp]
	\centering
		\includegraphics[width=0.40\textwidth]{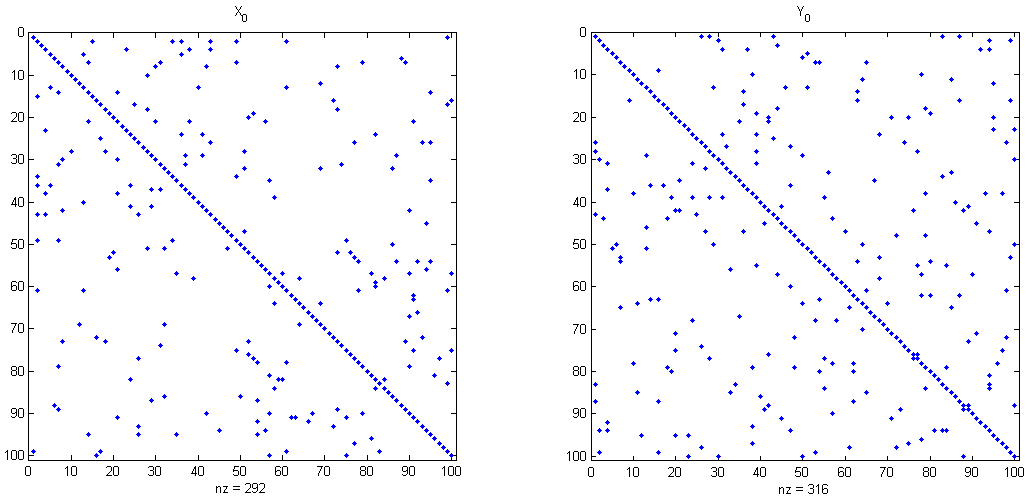}
	\caption{Sparse Kronecker matrix representation. Left panel: left Kronecker factor. Right panel: right Kronecker factor. As the Kronecker-product covariance matrix is of dimension $10,000 \times 10,000$ standard Glasso is not practically implementable for this example.}
	\label{fig: sim4_matrices}
	\vfill
		 \includegraphics[width=0.40\textwidth]{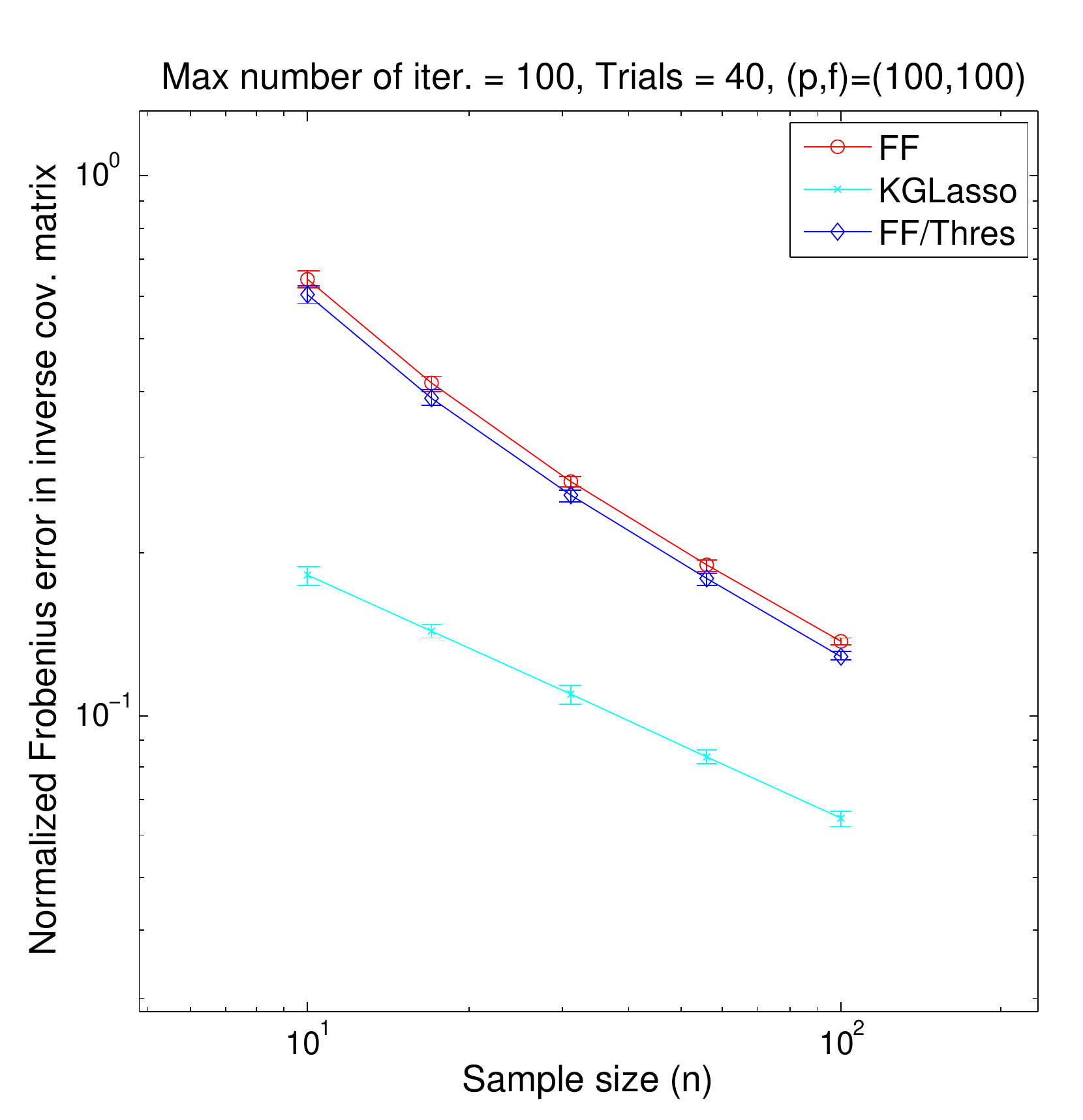}
	\caption{Normalized RMSE performance for precision matrix as a function of sample size $n$. KGlasso (Kronecker graphical lasso) uniformly outperforms FF (flip-flop) algorithm and FF/Thres (flip-flop thresholding) for all $n$. Here, $p=f=100$ and $N_{MC}=40$. The error bars are centered around the mean with $\pm$ one standard deviation. For $n=10$, there is a $72 \%$ RMSE reduction from the FF to KGLasso solution and a $70 \%$ RMSE reduction from the FF/Thres to KGLasso.  }
	\label{fig: sim4_frob_inv}
	\vfill
		 \includegraphics[width=0.40\textwidth]{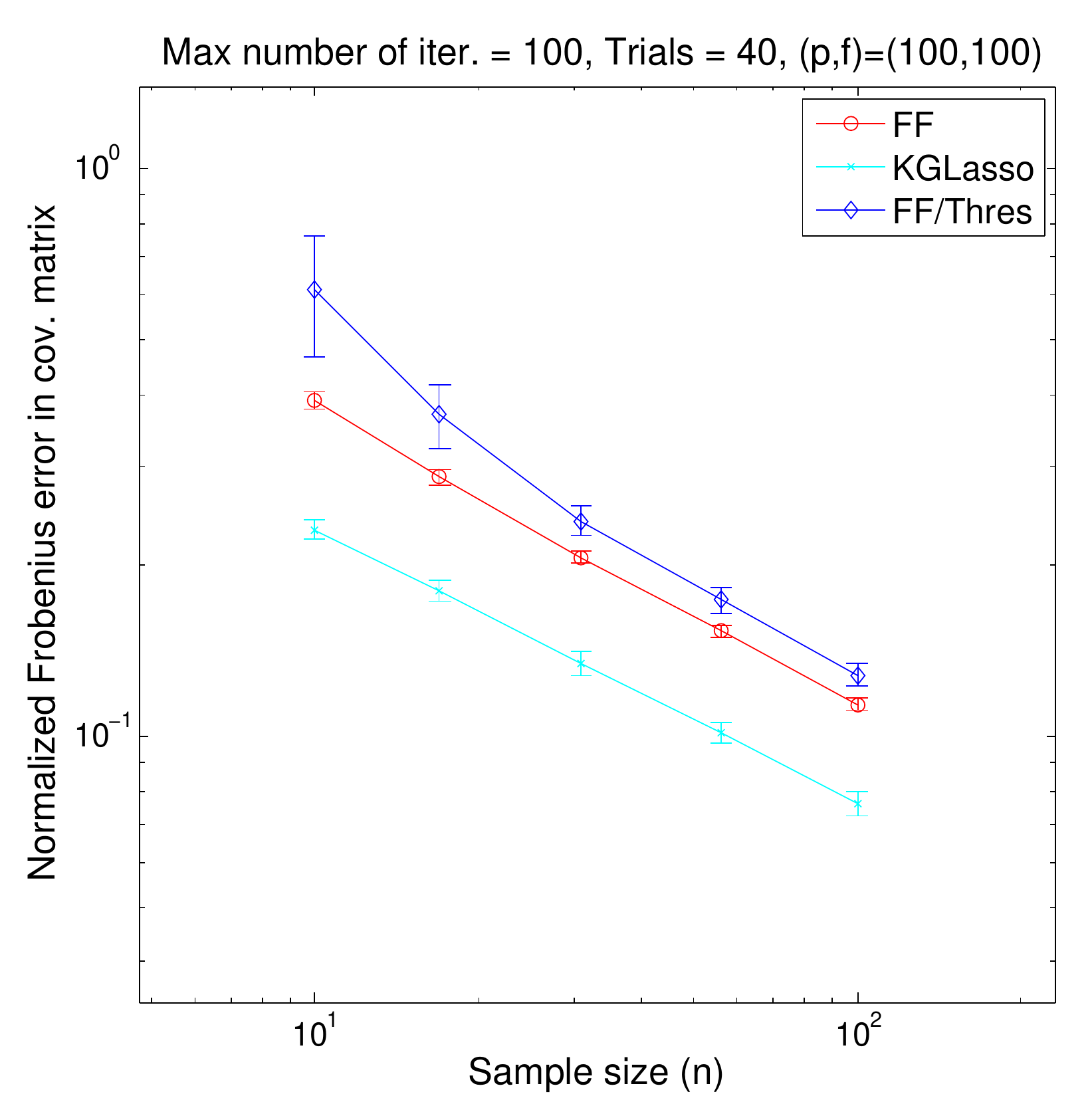}
	\caption{Normalized RMSE performance for covariance matrix as a function of sample size $n$. KGlasso (Kronecker graphical lasso) uniformly outperforms FF (flip-flop) algorithm for all $n$. Here, $p=f=100$ and $N_{MC}=40$. The error bars are centered around the mean with $\pm$ one standard deviation. For $n=10$, there is a $41 \%$ RMSE reduction from the FF to KGLasso solution and a $62 \%$ RMSE reduction from the FF/Thres to KGLasso. }
	\label{fig: sim4_frob_cov}
\end{figure}

We finally remark that the benefit obtained in the reduced convergence rate is not only due to the covariance estimation method chosen, but to the problem it addresses as well-i.e. the assumed true covariance structure.

\subsection{Empirical Rate Comparison}
\begin{figure}[htp]
	\centering
		\includegraphics[width=0.50\textwidth]{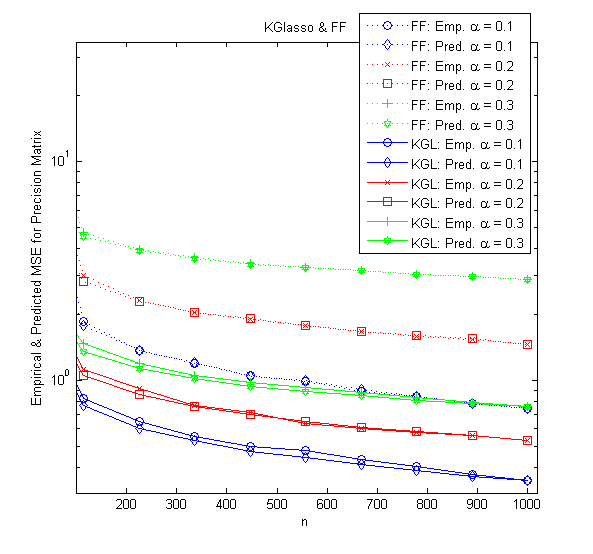}
	\caption{ Precision Matrix MSE convergence as a function of sample size $n$ for FF and KGlasso. The dimensions of the Kronecker factor matrices grow as a function of $n$ as: $p(n)=f(n)=\lceil 8 \cdot n^\alpha \rceil$. The true Kronecker factors were set to identity (so their inverses are fully sparse). The predicted MSE curves according to Thm. \ref{thm: FF_optimal_rate} and Thm. \ref{thm: KGL_optimal_rate} are also shown. For both KGlasso and FF, the predicted MSE matches the empirical MSE well, thus verifying the rate expressions (\ref{FF_rate_2}) and (\ref{KGL_perfect_rate}). }
	\label{fig: KGlasso_FF_curves}
\end{figure}

Next, we illustrate the rates obtained in for the dimension setting $p(n)=f(n)= \lceil 8 n^\alpha \rceil$, where $\alpha\in\{0.1,0.2,0.3\}$. According to the theory developed, for large $n$, the MSE converges to zero at a certain convergence rate. The predicted rates of FF and KGlasso are fitted on top of the empirical MSE curves by ensuring intersection at $n=1000$. Fig. \ref{fig: KGlasso_FF_curves} shows that the empirical rates match the predicted rates well.

\begin{figure}[htp]
	\centering
		\includegraphics[width=0.50\textwidth]{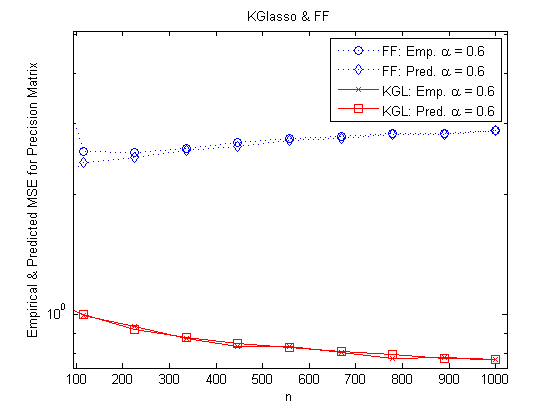}
	\caption{ Precision Matrix MSE as a function of sample size $n$ for FF and KGlasso. The dimensions of the Kronecker factor matrices grow as a function of $n$ as: $p(n)=f(n)=\lceil n^{0.6} \rceil$. The true Kronecker factors were set to identity (so their inverses are fully sparse). The predicted MSE curves according to Thm. \ref{thm: FF_optimal_rate} and Thm. \ref{thm: KGL_optimal_rate} are also shown. As predicted by our theory, and by the predicted convergent regions of $(n,p)$ for FF and KGlasso in Fig. \ref{fig: visRate},  the MSE of the FF diverges while the MSE of the KGlasso converges as $n$ increases. }
	\label{fig: KGlasso_FF_div_conv}
\end{figure}

We also show a borderline case $p=f=\lceil n^{0.6} \rceil$. In this case, according to Thm. \ref{thm: FF_optimal_rate} and Thm. \ref{thm: KGL_optimal_rate}, the FF diverges (MSE increases in $n$), while the KGlasso converges (MSE decreases in $n$). This is illustrated in Fig. \ref{fig: KGlasso_FF_div_conv}. Our predicted rates are plotted on top of the empirical curves.

\section{Conclusion}
We established high dimensional consistency for Kronecker Glasso algorithms that use iterative $\ell_1$-penalized likelihood optimization that exploit both Kronecker structure and sparsity of the covariance. A tight MSE convergence rate was derived for KGlasso, showing significantly better MSE performance than standard Glasso \cite{Rothman, ModelSel} and FF \cite{EstCovMatKron}. Simulations validated our theoretical predictions. 

As expected, the proposed KGlasso algorithm outperforms other algorithms (Glasso, FF) that do not exploit all prior knowledge about the covariance matrix, i.e., sparsity and Kronecker product structure, that KGlasso exploits. The theory and experiments in this paper establish that this performance gain is substantial, more so as the variable dimension increases. Furthermore, as compared to a simple thresholded FF algorithm, which does account for both sparsity and Kronecker structure, KGlasso has significantly better estimation performance.

\section*{Acknowledgement}
The authors thank Prof. Mark Rudelson for very helpful discussions on large deviation theory. The research reported in this paper was supported in part by ARO grant W911NF-11-1-0391.

\appendices

\section{Proof of Lemma \ref{dual_lemma}}
\begin{IEEEproof}
\begin{enumerate}
	\item Let $\theta \in (0,1)$. Let $\mathbf{X}_1,\mathbf{X}_2\in S_{++}^p$. Then, by the properties of the Kronecker product and trace:
	\begin{align*}
		\tr(&((\theta \mathbf{X}_1+ (1-\theta) \mathbf{X}_2) \otimes \mathbf{Y})\hat{\bS}_n) \\ 
			& \quad = \theta \tr((\mathbf{X}_1 \otimes \mathbf{Y})\hat{\bS}_n) + (1-\theta) \tr((\mathbf{X}_2 \otimes \mathbf{Y})\hat{\bS}_n)
	\end{align*}
	The function $g(\mathbf{X}_1):=-\log\det(\mathbf{X}_1)$ is a convex function in $\mathbf{X}_1$ over the set $S_{++}^p$ \cite{ConvexOpt}. By the triangle inequality:
	\begin{equation*}
		|\theta \mathbf{X}_1 + (1-\theta) \mathbf{X}_2 |_1 \leq \theta |\mathbf{X}_1|_1 + (1-\theta) |\mathbf{X}_2|_1
	\end{equation*}
	Finally, the sum of convex functions is convex.	The set $S_{++}^p$ is a convex set for any $p\in \mathbb{N}$. The other half of the argument follows by symmetry.
	\item By symmetry we only need prove that (\ref{dualX}) is the dual of $\min_{\mathbf{Y}\in S_{++}^f} J_\lambda(\bX,\bY)$. By standard duality relations between $\ell_1$ and $\ell_\infty$ norms \cite{ConvexOpt} and symmetry of $\bY$: 
	\begin{equation*}
		|\mathbf{Y}|_1=\max_{\mathbf{U}\in S^f: |\mathbf{U}|_{\infty}\leq 1}{\tr(\mathbf{Y}\mathbf{U})}
	\end{equation*}
	The maximum is attained at $\mathbf{U}_{i,j}= \frac{\mathbf{Y}_{i,j}}{|\mathbf{Y}_{i,j}|}$ for $\mathbf{Y}_{i,j}\neq 0$ and at $\mathbf{U}_{i,j}=0$ for $\mathbf{Y}_{i,j}=0$. Using this in (\ref{J_lambda_func}) and invoking the saddlepoint inequality:
	\begin{align}
	  &\min_{\mathbf{Y}\in S_{++}^f}  {\tr((\mathbf{X}\otimes \mathbf{Y})\hat{\bS}_n)-p\log\det(\mathbf{Y})+p\lambda_Y |\bY|_1} \nonumber \\
	  	&= \min_{\mathbf{Y}\in S_{++}^f} \max_{|\mathbf{U}|_{\infty}\leq \lambda_Y} \Big\{ \tr((\mathbf{X}\otimes \mathbf{Y})\hat{\bS}_n)-p\log\det(\mathbf{Y})\nonumber \\
	  	& \qquad +p \tr(\mathbf{Y}\mathbf{U}) \Big\} \nonumber \\
	    &\geq \max_{|\mathbf{U}|_{\infty}\leq \lambda_Y} \min_{\mathbf{Y}\in S_{++}^f} \Big\{ \tr((\mathbf{X}\otimes \mathbf{Y})\hat{\bS}_n)-p\log\det(\mathbf{Y}) \nonumber \\
	    & \qquad +p \tr(\mathbf{Y}\mathbf{U}) \Big\} \label{saddle_ptY}
	\end{align}
When the equality in (\ref{saddle_ptY}) is achieved, $(\bU,\mathbf{Y})$ is a saddlepoint and the duality gap is zero. Rewrite the objective function, denoted $\tilde{J}_\lambda(\cdot,\cdot)$, in the minimax operation (\ref{saddle_ptY}):
	\begin{equation*}
		\tilde{J}_\lambda(\bX,\bY):= \tr((\bX\otimes \bY)(\hat{\bS}_n+\tilde{\bU}(\bX)))-p\log\det(\bY)
	\end{equation*}
	where $\tilde{\mathbf{U}}(\mathbf{X})=p \frac{\bI_p \otimes \bU}{\tr(\bX)}$. Define $\bM=\hat{\bS}_n + \tilde{\bU}(\bX)$. To evaluate $\min_{\bY \in S_{++}^f}{\tilde{J}_\lambda(\bX,\bY)}$ in (\ref{saddle_ptY}), we invoke the KKT conditions to obtain the solution $\bY=\left( \frac{1}{p} \sum_{i,j=1}^p{\bX_{i,j} \bM(j,i))} \right)^{-1}$. Define $\bW=\bY^{-1}$ as the dual space variable. Using this in (\ref{saddle_ptY}):
	\begin{equation} \label{dualY_full}
		\max_{|\bW-\frac{1}{p}\sum_{i,j=1}^p{\bX_{i,j} \hat{\bS}_n(j,i)}|_{\infty}\leq \lambda_Y } { \{p \log\det(\mathbf{W}) + pf\} }
	\end{equation}
	where the constraint set was obtained in terms of $\bW$ by observing that $\tilde{\mathbf{U}}(\mathbf{X})(j,i) = \frac{p \mathbf{U}}{\tr(\mathbf{X})} I(j=i)$, and $I(\cdot)$ is the indicator function. 
	It is evident that (\ref{dualY_full}) is equivalent to (\ref{dualY}).

	\item It suffices to verify that the duality induced by the saddle point formulation is equivalent to Lagrangian duality (see Section 5.4 in \cite{ConvexOpt}). Slater's constraint qualification (see Section 5.3.2 in \cite{ConvexOpt}) trivially holds for the convex problem $\min_{\mathbf{Y} \in S_{++}^f}J_\lambda(\mathbf{X},\mathbf{Y})$ and and the corresponding convex problem $\min_{\mathbf{Y} \in S_{++}^f}\tilde{J}_\lambda(\mathbf{X},\mathbf{Y})$. Since the objective function of each dual problem has an optimal objective that is bounded below, Slater's constraint qualification also implies that the dual optimal solution is attained.

	\item From \cite{EstCovMatKron}, it follows that if $\hat{\bS}_n$ is p.d., each ``compression step'' (see lines 6 and 8 in Algorithm \ref{alg: algKGL}) yields a p.d. matrix. Combining this with the positive definiteness of the Glasso estimator \cite{ModelSel}, we conclude that the first subiteration of KGlasso yields a p.d. matrix. A simple induction, combined with the fact that the Kronecker product of p.d. matrices is p.d., establishes that (\ref{dualY}) and (\ref{dualX}) are p.d.

\end{enumerate}

\end{IEEEproof}

\section{Proof of Theorem \ref{convergence_fixed_point}}
\begin{IEEEproof}
	Recall that the basic optimization problem (\ref{opt_prob}) is
	\begin{equation*}
		\min_{\bX \in S_{++}^p, \bY \in S_{++}^f} J_\lambda(\bX,\bY)
	\end{equation*}
	Let $J^*:=\inf_{\mathbf{X} \in S_{++}^p, \mathbf{Y} \in S_{++}^f}{J_\lambda(\mathbf{X},\mathbf{Y})}$ be the optimal primal value. Note that $J_\lambda^* > -\infty$ when $n \geq \max(\frac{p}{f},\frac{f}{p})+1$. Now, consider the first step in Algorithm \ref{alg: algKGL}. Fix $\mathbf{X}=\mathbf{X}^{(k-1)}$ and optimize over $\mathbf{Y} \in S_{++}^f$. Invoking Lemma \ref{dual_lemma}, we have $\mathbf{Y}^{(k)}=\arg\min_{\mathbf{Y}\in S_{++}^f}{J_\lambda(\mathbf{X}^{(k-1)},\mathbf{Y})}$. Note, by induction $\mathbf{Y}^{(k)}$ remains positive definite if $\mathbf{X}^{(0)}$ is positive definite. Considering the second step in Algorithm \ref{alg: algKGL}, we fix $\mathbf{Y}=\mathbf{Y}^{(k)}$ and obtain $\mathbf{X}^{(k)}=\arg\min_{\mathbf{X}\in S_{++}^p}{J_\lambda(\mathbf{X},\mathbf{Y}^{(k)})}$, so that
	\begin{equation} \label{monotonicity}
		J_\lambda(\bX^{(k)},\bY^{(k)}) \leq	J_\lambda(\bX^{(k-1)},\bY^{(k)}) \leq J_\lambda(\bX^{(k-1)},\bY^{(k-1)})
	\end{equation}
	By induction on the number of iterations of the penalized flip-flop algorithm, we conclude that the iterates yield a nonincreasing sequence of objective functions. Since $\lambda_X |\bX|_1, \lambda_Y |\bY|_1 \geq 0$, we see that the objective function evaluated at the Kronecker structured MLE provides a lower bound to the optimal primal value \footnote{The Kronecker structured MLE $(\bX_{MLE},\bY_{MLE})$ exists for $n \geq \max(\frac{p}{f},\frac{f}{p})+1$.}
	\begin{equation} \label{LB_ML}
		J_\lambda(\bX_{KGlasso},\bY_{KGlasso}) \geq J_\lambda^* \geq J_\lambda(\bX_{MLE},\bY_{MLE}) > -\infty
	\end{equation}
	Thus, the sequence $\{J_\lambda^{(k)}:k\geq 0\}$ forms a nonincreasing sequence bounded below (since for $n>pf$, the log-likelihood function is bounded above by the log-likelihood evaluated at the sample mean and sample covariance matrix). The monotone convergence theorem for sequences \cite{IntroRealAnalysis} implies that $\{J_\lambda^{(k)}\}$ converges monotonically to $J_\lambda^{(\infty)}=\inf_k{J_\lambda^{(k)}}$. By the alternating minimization, we conclude that the sequence of iterates $\{(\mathbf{X}^{(k)},\mathbf{Y}^{(k)})\}_k$ converges since the minimizer at each Glasso step is unique.
\end{IEEEproof}

\section{Subdifferential Calculus Review}
As sparse Kronecker Glasso involves non-smooth objective functions, we review a few definitions and facts from subdifferential calculus \cite{VariationalAnalysis}.

\begin{definition} \label{def_attentive}
	By \text{J-attentive} convergence denoted as, $\mathbf{x}^n \stackrel{J}{\rightarrow} \mathbf{x}$, we mean that: $\mathbf{x}^n\to \mathbf{x}$ with $J(\mathbf{x}^n)\to J(\mathbf{x})$ as $n\to \infty$.
\end{definition}
The role of \text{J-attentive} convergence is to make sure that subgradients at a point $\overline{\mathbf{x}}$ reflect no more than the local geometry of $epi(J)$ around $(\overline{\mathbf{x}},J(\overline{\mathbf{x}}))$.

\begin{definition} \label{def_subgradient}
Consider a proper lower semicontinuous (LSC) function $g:\RR^d\to \RR \cup \{+\infty\}$. Let $\overline{\mathbf{x}}$ be such that $J(\overline{\mathbf{x}})<\infty$. \\
For $\mathbf{v}\in \RR^d$, \\
a) $\mathbf{v}$ is a \textit{regular subgradient} of $J$ at $\overline{\mathbf{x}}$ (i.e., $\mathbf{v} \in \hat{\partial}J(\overline{\mathbf{x}})$) if $\liminf_{\mathbf{x}\neq \overline{\mathbf{x}}, \mathbf{x}\to \overline{\mathbf{x}}}{\frac{J(\mathbf{x})-J(\overline{\mathbf{x}})-\mathbf{v}^T(\mathbf{x}-\overline{\mathbf{x}})}{\nn \mathbf{x} - \overline{\mathbf{x}}\nn}} \geq 0$. \\
b) $\mathbf{v}$ is a \textit{general subgradient} of $J$ at $\overline{\mathbf{x}}$ (i.e., $\mathbf{v} \in \partial J(\overline{\mathbf{x}})$) if there exists subsequences $\mathbf{x}^n \stackrel{J}{\rightarrow} \overline{\mathbf{x}}$ and $\mathbf{v}^n \in \hat{\partial} J(\mathbf{x}^n)$ such that $\mathbf{v}^n \to \mathbf{v}$.
\end{definition}

Let $\overline{\mathbf{x}}$ be such that $J(\overline{\mathbf{x}})<\infty$. It can be shown that $\partial J(\overline{\mathbf{x}}) = \limsup_{\mathbf{x} \stackrel{J}{\rightarrow}\overline{\mathbf{x}}}{\hat{\partial}J(\mathbf{x})}$, $\hat{\partial}J(\overline{\mathbf{x}}) \subset \partial J(\overline{\mathbf{x}})$ and both sets are closed.

Define the set of critical points $C_J:=\{\mathbf{x}: 0\in \partial J(\mathbf{x})\}=C_{J,min} \cup C_{J,saddle} \cup C_{J,max}$, where $C_{J,min}$ contains all the local minima, $C_{J,saddle}$ contains all the saddle points and $C_{J,max}$ contains all the local maxima.

\begin{definition}
	Let $A \subseteq \RR^n$. Define the distance from a point $\mathbf{x} \in \RR^n$ to the set $A$ as $d(\mathbf{x},A):=\inf_{\mathbf{a}\in A}{\nn \mathbf{x}-\mathbf{a} \nn_2}$.
\end{definition}

\section{Properties of objective function $J_\lambda$}
The following set of properties will be used in Lemmas \ref{lemma: subdifferential}, \ref{prelim_lemma} and Theorem \ref{convergence_theorem_critical}.

\begin{property} \label{property1}
\indent 1. $J_0:\RR^d \to \RR$ is continuously differentiable (i.e., $f_0 \in C^1$) \\
\indent 2. $\nabla J_0:\RR^d \to \RR^d$ is uniformly continuous on bounded subsets $B \subset \RR^d$ \\
\indent 3. $J_i:\RR^{d_i}\to \RR \cup \{+\infty\}$ is proper \footnote{A function $J:\mathbb{X} \to \RR \cup \{\pm \infty\}$ is proper if $dom(J)=\{x\in \mathbb{X}: J(x) < \infty\} \neq \emptyset$ and $J(x)>-\infty, \forall x \in \mathbb{X}$. } and lower semicontinuous (LSC), for $i=1,\dots,k$ \\
\indent 4. $\eta_i:\RR^d\to\RR_{+}$ is uniformly continuous and bounded on bounded subsets $B \subset \RR^d$, for $i=1,\dots,k$ \\
\indent 5. $J_\lambda$ is bounded below-i.e. $J_\lambda^*>-\infty$ \\
\indent 6. $J_\lambda$ is strictly convex in at least one block (for all the rest of the blocks held fixed) 
\end{property}
where $J_\lambda^*=\inf_{\mathbf{X}_i \in S_{++}^{d_i}}{J_\lambda(\mathbf{X}_1,\dots,\mathbf{X}_k)}$ is the optimal primal value.

\section{Lemma \ref{lemma: subdifferential}}
\begin{lemma} \label{lemma: subdifferential}
	Given the notation established in Definition \ref{def_subgradient} and $J_\lambda$ given by (\ref{general_f}), we have:
	\begin{align}
		\partial J_\lambda(\mathbf{x}_1,\dots,\mathbf{x}_k) &= \times_{i=1}^k \{\nabla_{\mathbf{x}_i}J_0(\mathbf{x}_1,\dots,\mathbf{x}_k)+\partial J_i(\mathbf{x}_i)\nonumber \\
			&\qquad +\bar{\lambda}_i \partial \eta_i(\mathbf{x}_i)\} \nonumber \\
			&= \times_{i=1}^{k} \{\partial_{\mathbf{x}_i}J_\lambda(\mathbf{x}_1,\dots,\mathbf{x}_k)\}  \label{d_general_f}
	\end{align}
	where $\partial_{\mathbf{x}_i}J_\lambda(\mathbf{x}_1,\dots,\mathbf{x}_k)$ is the partial differential operator while all $\{\mathbf{x}_j:j\neq i\}$ are held fixed.
\end{lemma}
\begin{IEEEproof}
	First note that we have:
	\begin{align}
		\partial & J_\lambda(\mathbf{x}_1,\dots,\mathbf{x}_k) = \nabla J_0(\mathbf{x}_1,\dots,\mathbf{x}_k) + \partial\{\sum_{i=1}^k{J_i(\mathbf{x}_i)} \nonumber \\
			&\qquad +  \sum_{i=1}^k{\bar{\lambda}_i \eta_i(\mathbf{x}_i)}\} \label{e1} \\
			&= \nabla J_0(\mathbf{x}_1,\dots,\mathbf{x}_k) + \partial\{\sum_{i=1}^k{J_i(\mathbf{x}_i)}\} + \partial\{\sum_{i=1}^k{\bar{\lambda}_i \eta_i(\mathbf{x}_i)}\} \label{e2} \\
			&= \nabla J_0(\mathbf{x}_1,\dots,\mathbf{x}_k) + \times_{i=1}^k \{\partial J_i(\mathbf{x}_i)\} + \times_{i=1}^k\{\bar{\lambda}_i \partial \eta_i(\mathbf{x}_i) \} \label{e3} \\
			&= \times_{i=1}^k\{\nabla_{\mathbf{x}_k} J_0(\mathbf{x}_1,\dots,\mathbf{x}_k) + \partial J_i(\mathbf{x}_i) + \bar{\lambda}_i \partial \eta_i(\mathbf{x}_i)\} \label{e4}
	\end{align}
	where (\ref{e1}) follows from Property \ref{property1} and Exercise 8.8(c) in \cite{VariationalAnalysis}, (\ref{e2}) follows from Corollary 10.9 in \cite{VariationalAnalysis}, (\ref{e3}) follows from Proposition 10.5 and Equation 10(6) p.438 in \cite{VariationalAnalysis} since $\lambda_i>0$, and finally (\ref{e4}) follows from Minkowski sum properties.
	
\end{IEEEproof}

\section{Lemma \ref{prelim_lemma}}
\begin{lemma} \label{prelim_lemma}
	Let $m$ denote the iteration index. For $m \in \NN$, define:
	\begin{align*}
		(\mathbf{x}_1^m)^\circ & := \nabla_{\mathbf{x}_1}J_0(\mathbf{x}_1^m,\mathbf{x}_2^m\dots,\mathbf{x}_k^m) \\
			&\quad - \nabla_{\mathbf{x}_1}J_0(\mathbf{x}_1^m,\mathbf{x}_2^{m-1}\dots,\mathbf{x}_k^{m-1}) \\
		(\mathbf{x}_2^m)^\circ & := \nabla_{\mathbf{x}_2}J_0(\mathbf{x}_1^m,\mathbf{x}_2^m\dots,\mathbf{x}_k^m) \\
			&\quad - \nabla_{\mathbf{x}_2}J_0(\mathbf{x}_1^m,\mathbf{x}_2^m,\mathbf{x}_{3}^{m-1}\dots,\mathbf{x}_k^{m-1}) \\
		&\vdots \\
		(\mathbf{x}_j^m)^\circ & := \nabla_{\mathbf{x}_j}J_0(\mathbf{x}_1^m,\mathbf{x}_2^m\dots,\mathbf{x}_k^m) \\
			&\quad - \nabla_{\mathbf{x}_j}J_0(\mathbf{x}_1^m,\dots,\mathbf{x}_j^m,\mathbf{x}_{j+1}^{m-1}\dots,\mathbf{x}_k^{m-1}) \\
		&\vdots \\
		(\mathbf{x}_k^m)^\circ &:= 0
	\end{align*}
	Then, $((\mathbf{x}_1^m)^\circ,\dots,(\mathbf{x}_k^m)^\circ) \in \partial J_\lambda(\mathbf{x}_1^m,\dots,\mathbf{x}_k^m)$. Also, for all convergent subsequences $(\mathbf{x}^{m_j})_j$ of the sequence $(\mathbf{x}^m)_m$, we have 
	\begin{equation} \nonumber
		d(0, \partial J_\lambda(\mathbf{x}_1^{m_j}, \dots, \mathbf{x}_k^{m_j})) \to 0 \text{ as } j\to \infty
	\end{equation}
\end{lemma}
\begin{IEEEproof}
From Algorithm \ref{alg: alg2}, we have:
	\begin{align*}
		\mathbf{x}_1^m &\in \arg \min_{\mathbf{x}_1} {J_\lambda(\mathbf{x}_1,\mathbf{x}_2^{m-1},\dots,\mathbf{x}_k^{m-1})} \\
		\mathbf{x}_2^m &\in \arg \min_{\mathbf{x}_2} {J_\lambda(\mathbf{x}_1^m,\mathbf{x}_2,\mathbf{x}_3^{m-1},\dots,\mathbf{x}_k^{m-1})} \\
		&\vdots \\
		\mathbf{x}_k^m &\in \arg \min_{\mathbf{x}_k} {J_\lambda(\mathbf{x}_1^m,\dots,\mathbf{x}_{k-1}^m,\mathbf{x}_k)}
	\end{align*}
	The first subiteration step of the algorithm implies that $0\in \partial_{\mathbf{x}_1} J_\lambda(\mathbf{x}_1^m,\mathbf{x}_2^{m-1},\dots,\mathbf{x}_k^{m-1})$, the second subiteration step implies $0\in \partial_{\mathbf{x}_2} J_\lambda(\mathbf{x}_1^m,\mathbf{x}_2^{m},\mathbf{x}_3^{m-1},\dots,\mathbf{x}_k^{m-1})$, etc. Rewriting these using Lemma \ref{lemma: subdifferential}, we have:
	\begin{align*}
	 	0 &\in \nabla_{\mathbf{x}_1}J_0(\mathbf{x}_1^m,\mathbf{x}_2^{m-1},\dots,\mathbf{x}_k^{m-1}) + \partial J_1(\mathbf{x}_1^m) + \bar{\lambda}_1 \partial \eta_1(\mathbf{x}_1^m) \\
	 	0 &\in \nabla_{\mathbf{x}_2}J_0(\mathbf{x}_1^m,\mathbf{x}_2^m,\mathbf{x}_3^{m-1},\dots,\mathbf{x}_k^{m-1}) + \partial J_2(\mathbf{x}_2^m) + \bar{\lambda}_2 \eta_2(\mathbf{x}_2^m) \\
	 		&\vdots \\
	 	0 &\in \nabla_{\mathbf{x}_k}J_0(\mathbf{x}_1^m,\mathbf{x}_2^m,\dots,\mathbf{x}_k^m) + \partial J_k(\mathbf{x}_k^m) + \bar{\lambda}_k \partial \eta_k(\mathbf{x}_k^m)
	\end{align*}
This implies that for $i=1,\dots,k$:
	\begin{equation*}
		(\mathbf{x}_i^m)^\circ \in \nabla_{\mathbf{x}_i}J_0(\mathbf{x}_1^m,\mathbf{x}_2^m,\dots,\mathbf{x}_k^m) + \partial J_i(\mathbf{x}_i^m) + \bar{\lambda}_i \partial \eta_i(\mathbf{x}_i^m)
	\end{equation*}
	It is important to note that $\partial \eta_i(\mathbf{x}) \neq \emptyset, \forall \mathbf{x} \in \RR^{d_i}$, for $i=1,\dots,k$, as a result of property \ref{property1}.4. To see why, apply Corollary 8.10 in \cite{VariationalAnalysis} since $\eta_i$ is finite and locally LSC at every point in its domain. This in turn implies $((\mathbf{x}_1^m)^\circ,\dots,(\mathbf{x}_k^m)^\circ) \in \partial J_\lambda(\mathbf{x}_1^m,\dots,\mathbf{x}_k^m)$ by Lemma \ref{prelim_lemma}.

Now, take an arbitrary convergent subsequence $(\mathbf{x}_1^{m_j},\dots,\mathbf{x}_k^{m_j})_j$ of $(\mathbf{x}_1^m,\dots,\mathbf{x}_k^m)_m$. The convergence of $(\mathbf{x}_1^{m_j},\dots,\mathbf{x}_k^{m_j})_j$ implies the convergence of $(\mathbf{x}_1^{m_j},\mathbf{x}_2^{m_j-1},\dots,\mathbf{x}_k^{m_j-1})_j$, and  $(\mathbf{x}_1^{m_j},\dots,\mathbf{x}_i^{m_j},\mathbf{x}_{i+1}^{m_j-1},\dots,\mathbf{x}_k^{m_j-1})_j$ for $i=2,\dots,k-1$. Taking $j\to \infty$ and using properties \ref{property1}.2, we see that $\lim_{j\to\infty} {d(0,\partial J_\lambda(\mathbf{x}_1^{m_j},\dots,\mathbf{x}_k^{m_j}))} = 0$ since $\lim_{j\to\infty}{((\mathbf{x}_1^{m_j})^\circ,\dots,(\mathbf{x}_k^{m_j})^\circ)}=(0,\dots,0)$.

\end{IEEEproof}

\section{Proof of Theorem \ref{convergence_theorem_critical}}
\begin{IEEEproof}

\begin{enumerate}
	\item Let $L(\mathbf{x}^0)=L(\mathbf{x}_1^0,\dots,\mathbf{x}_k^0)$ be the set of all limit points of $(\mathbf{x}^m)_{m\geq 0}$ starting from $\mathbf{x}^0$. The block-coordinate descent algorithm, Algorithm \ref{alg: alg2}, implies
	\begin{align*}
		J_0&(\mathbf{x}_1^m,\mathbf{x}_2^{m-1},\dots,\mathbf{x}_k^{m-1}) + J_1(\mathbf{x}_1^m) + \bar{\lambda}_1 \eta_1(\mathbf{x}_1^m) \\
			&\leq J_0(\alpha_1,\mathbf{x}_2^{m-1},\dots,\mathbf{x}_k^{m-1}) + J_1(\mathbf{\alpha}_1) + \bar{\lambda}_1 \eta_1(\mathbf{\alpha}_1)
	\end{align*}
	for any $\mathbf{\alpha}_1 \in \RR^{d_1^2}$. Now, assume there exists a subsequence $(\mathbf{x}^{m_j})_j$ of $(\mathbf{x}^m)_m$ that converges to $\mathbf{x}^*$, where $\mathbf{x}^*$ is a limit point. This implies that $(\mathbf{x}_1^{m_j},\mathbf{x}_2^{m_j-1},\dots,\mathbf{x}_k^{m_j-1}) \to \mathbf{x}^*$ as $j\to\infty$. The above inequality combined with properties \ref{property1}.1 and \ref{property1}.4 (i.e. the continuity $J_0$ and $\eta_i$) then implies that
	\begin{align*}
		\limsup_{j\to\infty} &{J_1(\mathbf{x}_1^{m_j})} + J_0(\mathbf{x}_1^*,\dots,\mathbf{x}_k^*) \leq J_1(\mathbf{\alpha}_1) \\
			&\quad + J_0(\mathbf{\alpha}_1,\mathbf{x}_2^*,\dots,\mathbf{x}_k^*) + \bar{\lambda}_1 (\eta_1(\mathbf{\alpha}_1) - \eta_1(\mathbf{x}_1^*))
	\end{align*}
	for all $\mathbf{\alpha}_1 \in \RR^{d_1^2}$. Taking $\mathbf{\alpha}_1 = \mathbf{x}_1^*$ then yields $\limsup_{j\to \infty}{J_1(\mathbf{x}_1^{m_j})} \leq J_1(\mathbf{x}_1^*)$. Using the lower semicontinuity property of $J_1$ (property \ref{property1}.3), we have $\liminf_{j\to\infty}{J_1(\mathbf{x}_1^{m_j})} \geq J_1(\mathbf{x}_1^*)$. Thus, $\lim_{j\to\infty} {J_1(\mathbf{x}_1^{m_j}) = J_1(\mathbf{x}_1^*)}$.
	
	By a similar line of reasoning, it can be shown that $J_i(\mathbf{x}_i^{m_j}) \to J_i(\mathbf{x}_i^*)$ as $j \to \infty$, for $i=1,\dots,k$. As a result, $\sum_{i=1}^k{J_i(\mathbf{x}_i^{m_j})} \to \sum_{i=1}^k{J_i(\mathbf{x}_i^*)}$ as $j \to \infty$. Since $J_0(\cdot)$ is jointly continuous, $J_0(\mathbf{x}_1^{m_j},\dots,\mathbf{x}_k^{m_j}) \to J_0(\mathbf{x}_1^*,\dots,\mathbf{x}_k^*)$. By continuity of $\eta_i(\cdot)$, $\sum_{i=1}^k{\bar{\lambda}_i \eta_i(\mathbf{x}_i^{m_j})} \to \sum_{i=1}^k \bar{\lambda}_i \eta_i(\mathbf{x}_i^*)$. Thus, $J_\lambda(\mathbf{x}^{m_j}) \to J_\lambda(\mathbf{x}^*)$ as $j\to\infty$.
	
	Now, Lemma \ref{prelim_lemma} implies that $((\mathbf{x}^{m_j})^{\circ}) \in \partial J_\lambda(\mathbf{x}^{m^j})$. Since the subsequence $(\mathbf{x}^{m_j})_j$ is convergent, by Lemma \ref{prelim_lemma}, we have $(\mathbf{x}^{m_j})^\circ \to 0$ as $j \to \infty$. As a result, since $\partial J_\lambda(\mathbf{x}^{m_j})$ is closed (see Theorem 8.6 in \cite{VariationalAnalysis}) for all $j$, we conclude that $\mathbf{x}^* \in C_J$. Thus, $L(\mathbf{x}^0) \subseteq C_J$.
	
	We have thus proved that limit points are critical points of the objective function. 
	
	We can rule out convergence to local maxima thanks to property \ref{property1}.6. Let us show this rigorously. Assume there exists a local maximum at $\bx'=(\bx_1',\dots,\bx_k')$. Then, there exists $r>0$ such that $J_\lambda(\bx) \leq J_\lambda(\bx')$ for all $\bx$ such that $\nn \bx - \bx'\nn_2 < r$. Fix $\bx_i=\bx_i'$ for all $i\neq 1$. Without loss of generality, assume $J_\lambda$ is strictly convex in the first block. Since strict convexity is maintained through linear transformation, without loss of generality, assume $d_1=1$. Let $\epsilon<r$. Define $x_{1,\epsilon}=x_1' - \epsilon$ and $x_{2,\epsilon}=x_1' + \epsilon$. Define $x_\theta = \theta x_{1,\epsilon} + (1-\theta) x_{2,\epsilon}$, where $\theta \in (0,1)$. Since $\nn [x_\theta; \bx_{\neq 1}]-\bx' \nn_2 = |x_\theta - x_1'| = \epsilon (1-2\theta) < r$, by the local maximum definition, there exists $\epsilon\in(0,r)$ small enough such that
	\begin{equation*}
		\theta J_\lambda(x_{1,\epsilon},\bx_{\neq 1}') + (1-\theta) J_\lambda(x_{2,\epsilon},\bx_{\neq 1}') \leq J_\lambda(x_\theta,\bx_{\neq 1}')
	\end{equation*}
	for some $\theta\in (0,1)$. Since $\epsilon>0$, we have $x_{1,\epsilon} \neq x_{2,\epsilon}$, and this contradicts strict convexity. 
	Thus, there are no local maxima. \footnote{An alternative way to get a contradiction is to assume there exists a strict local maximum and use only convexity, instead of strict convexity.}
	
	Next, we use the non-existence of local maxima and continuity of $J_\lambda$ to rule out convergence to saddle points. Assume there exists a saddlepoint at $\bx_s$. Then, by definition, $0\in J_\lambda(\bx_s)$ and $\bx_s$ is not a local maximum or a local minimum. Since $\bx_s$ is not a local minimum, for all $\epsilon>0$, there exists a point $\bx'$ such that $\nn \bx'-\bx_s \nn_2<\epsilon$ and $J_\lambda(\bx_s)>J_\lambda(\bx')$. By continuity, it follows that there exists $\delta>0$ such that for all $\bx$ satisfying $\nn \bx-\bx'\nn_2<\delta$, we have $J_\lambda(\bx_s)>J_\lambda(\bx)$, which implies that $\bx_s$ is a local maximum. This is a contradiction and thus, $\bx_s$ is a local minimum. So, no saddle points exist.

	Theorem \ref{convergence_fixed_point} implies that $L(\mathbf{x}^0)$ is nonempty and singleton.
	
	\item We show that if we do not start at a local minimum, strict descent follows. Let $\mu(\cdot)$ denote the point-to-point mapping during one iteration step, i.e., $\mathbf{x}^{m+1}=\mu(\mathbf{x}^m)$. We show that if $\mathbf{x}^0 \notin C_J$, then $L(\mathbf{x}^0) \subseteq C_{J,min}$. The result then follows by using the proof of the first part \footnote{The first part of the proof showed $C_J = C_{J,min}$.}. To this end, let $\mathbf{x}^{'}$ be a fixed point under $\mu$, i.e., $\mu(\mathbf{x}^{'})=\mathbf{x}^{'}$. Then, the subiteration steps of the algorithm yield $0\in \partial_{\mathbf{x}_i} J_\lambda(\mathbf{x}_1^{'},\dots,\mathbf{x}_k^{'})$ for $i=1,\dots,k$, which implies $0\in \partial J_\lambda(\mathbf{x}^{'})$, i.e., $\mathbf{x}^{'}\in C_J$. The contrapositive implies that if $\mathbf{x}\notin C_J$, then $J_\lambda(\mu(\mathbf{x}))<J_\lambda(\mathbf{x})$ (strict descent). A simple induction on the number of iterations then concludes the proof.
\end{enumerate}
	
\end{IEEEproof}


\section{Lemma \ref{Isserlis_bound}}
The following technical lemma will be used in the proof of Lemma \ref{lemma: large_dev_optimal}.
\begin{lemma} \label{Isserlis_bound}
	Let $\bz \sim N(\mathbf{0},\bA_0\otimes \bB_0)$,where $\bA_0\in S_{++}^p, \bB_0\in S_{++}^f$. Then, for $m\geq 0$, we have the moment bound:
	\begin{align*}
		\E & \left[ \left( \sum_{i,j=1}^p \bX_{i,j} \left([\bz]_{(i-1)f+k}[\bz]_{(j-1)f+l}-[\bA_0]_{i,j}[\bB_0]_{k,l}  \right) \right)^{m+2} \right] \\
			&\leq (2m+2)!! p \left( \max_{1\leq k\leq f} [\bB_0]_{k,k} \nn\bX\nn_2 \nn\bA_0\nn_2 \right)^{m+2}
	\end{align*}
\end{lemma}

\begin{remark}
In the symmetric $\bX\in S^p$ case, the bound in Lemma \ref{Isserlis_bound} can be tightened to
\begin{align*}
		\E & \left[ \left( \sum_{i,j=1}^p \bX_{i,j} \left([\bz]_{(i-1)f+k}[\bz]_{(j-1)f+l}-[\bA_0]_{i,j}[\bB_0]_{k,l}  \right) \right)^{m+2} \right] \\
			&\leq (2m+2)!! (\max_{1\leq k\leq f} [\bB_0]_{k,k})^{m+2} \tr((\bX \bA_0)^{m+2})
\end{align*}
\end{remark}

\begin{IEEEproof}

Consider the index set $\{\{i_1,j_1\},\{i_2,j_2\},\dots,\{i_{m+2},j_{m+2}\}\}$. Define groups $G_k=\{i_k,j_k\}$ for $k=1,\dots,m+2$. Let the generic notation $\pi(\cdot)$ denote the permutation operator of a set of indices.

Define the set of indices $M_{m+2}=M_{m+2}(i_1,j_1,\dots,i_{m+2},j_{m+2})$ as the set containing sequences $(I_1,J_1,\dots,I_{m+2},J_{m+2})$ satisfying the properties:
\begin{enumerate}
	\item $\{I_1,J_1,\dots,I_{m+2},J_{m+2}\}$ is a permutation of the index set $\{i_1,j_1,\dots,i_{m+2},j_{m+2}\}$ \\
		-i.e. $\{I_1,J_1,\dots,I_{m+2},J_{m+2}\} = \pi(\{i_1,j_1,\dots,i_{m+2},j_{m+2}\})$
	\item For each $q\in\{1,\dots,m+2\}$, indices $I_q$ and $J_q$
		must belong to disjoint groups $\{G_k\}_{k=1}^{m+2}$
	\item Suppose a sequence $\{I_1,J_1,\dots,I_{m+2},J_{m+2}\}$ satisfies
		the first two properties. Then, add it to $M_{m+2}$ and $M_{m+2}$ does not
		contain (block-permuted) sequences of the form \\
		$\{\pi(\{\pi(\{I_1,J_1\}),\pi(\{I_2,J_2\}),\dots,\pi(\{I_{m+2},J_{m+2}\})\})\}$
\end{enumerate}
It can be shown that $\card(M_{m+2})=(2m+2)!!$.

As an illustrative example, consider the case $m=1$. 
\begin{example}
	For $m=1$, the set $M_{m+2}$ contains the following $4!!=8$ elements:
	\begin{align*}
		&\{\{i_1,i_2\},\{j_1,i_3\},\{j_2,j_3\}\}, 		\{\{i_1,i_2\},\{j_1,j_3\},\{j_2,i_3\}\}, \\
		&\{\{i_1,j_2\},\{j_1,i_3\},\{i_2,j_3\}\}, 		\{\{i_1,j_2\},\{j_1,j_3\},\{i_2,i_3\}\}, \\
		&\{\{i_1,i_3\},\{j_1,i_2\},\{j_2,j_3\}\}, 		\{\{i_1,i_3\},\{j_1,j_2\},\{i_2,j_3\}\}, \\
		&\{\{i_1,j_3\},\{j_1,j_2\},\{i_2,i_3\}\}, 		\{\{i_1,j_3\},\{j_1,i_2\},\{j_2,i_3\}\}.
	\end{align*}
	Of course, other equivalent possibilities for $M_{m+2}$ are possible.
\end{example}

Note that $\tr((\bX\bA_0)^{m+2})\geq 0$ for all $m\geq 0$. From Isserlis' formula \cite{Isserlis}, we have:
\begin{align*}
	& \E\left[ \left( \sum_{i,j=1}^p \bX_{i,j} \left([\bz]_{(i-1)f+k}[\bz]_{(j-1)f+l}-[\bA_0]_{i,j}[\bB_0]_{k,l}  \right) \right)^{m+2} \right] \\
	&= \sum_{i_1,j_1=1}^p \cdots \sum_{i_{m+2},j_{m+2}=1}^p \bX_{i_1,j_1} \cdots \bX_{i_{m+2},j_{m+2}} \\
			&\times  \E\Big[ \prod_{\alpha=1}^{m+2} \Big( [\bz]_{(i_\alpha-1)f+k}[\bz]_{(j_\alpha-1)f+l} -[\bA_0]_{i_\alpha,j_\alpha}[\bB_0]_{k,l} \Big) \Big] \\
			&\leq (\max_{1\leq k\leq f} [\bB_0]_{k,k})^{m+2} \sum_{i_1,j_1=1}^p \cdots \sum_{i_{m+2},j_{m+2}=1}^p \bX_{i_1,j_1} \cdots \bX_{i_{m+2},j_{m+2}}  \\
			&\quad \times \sum_{\{I_q,J_q\}_{q=1}^{m+2} \in M_{m+2}} \prod_{q=1}^{m+2} [\bA_0]_{I_q,J_q} \\
			&\leq (\max_{1\leq k\leq f} [\bB_0]_{k,k})^{m+2} (2m+2)!! p (\nn\bX \nn_2 \nn\bA_0 \nn_2)^{m+2}
\end{align*}

\end{IEEEproof}

\section{Lemma \ref{lemma: large_dev_optimal}}
The following lemma will be used in the proof of Theorem \ref{thm: FF_optimal_rate} and Theorem \ref{thm: KGL_optimal_rate}. The method of proof is by moment generating functions. A similar bound can be obtained under the same set of assumptions using standard decoupling arguments and Gaussian chaos Talagrand-based bounds. 
\begin{lemma} \label{lemma: large_dev_optimal}
	Let $\bX$ be a $p\times p$ data-independent matrix. Define the linear operator $\bT$ as $\bT(\bX) = \hat{\bB}(\bX^{-1})$, where $\hat{\bB}(\cdot)$ is defined in (\ref{B_update}). Assume $\max_{k}[\bB_0]_{k,k}, \nn \bX\nn_2, \nn \bA_0\nn_2$ are uniformly bounded constants as $p,f\to\infty$. Define $\bB_*:=\frac{\tr(\bX \bA_0)}{p} \bB_0$. Let $c,\tau>0$. Define $\psi(u)=\sum_{m=0}^\infty \frac{(2m+2)!!}{m!} u^m$ \footnote{The double factorial notation is defined as 

\[
m!! = \left\{ 
  \begin{array}{l l}
    m\cdot (m-2) \cdot \cdots \cdot 3 \cdot 1 & \quad \text{if $m>0$ is odd}\\
    m\cdot (m-2) \cdot \cdots \cdot 4 \cdot 2 & \quad \text{if $m>0$ is even}\\
    1 & \quad \text{if $m=-1$ or $m=0$}\\
  \end{array} \right.
\]  
	. }. Let $\bar{C} := \frac{4 (2+\tau)^2 \max(2,c) }{\psi(\frac{1}{2+\tau})}<\frac{np}{\log(\max(f,n))}$ \footnote{If $p=f=n^{c'}$ for some $c'>0$, this condition will hold for $n$ large enough.}. Then, with probability $1-\frac{2}{\max(f,n)^c}$,
	\begin{equation*}
		|\bT(\bX)-\bB_*|_\infty \leq \overline{k} \cdot \sqrt{4 \psi(\frac{1}{2+\tau}) \max(2,c)} \sqrt{\frac{\log(\max(f,n))}{np}}
	\end{equation*}
	where $\overline{k}=\max_{k}[\bB_0]_{k,k} \cdot \nn \bX\nn_2 \nn \bA_0\nn_2$.
\end{lemma}

\begin{remark}
	Choosing $c\leq 2$ in Lemma \ref{lemma: large_dev_optimal}, the best relative constant is obtained by taking $\tau$ to infinity, which yields $\sqrt{4 \psi(\frac{1}{2+\tau}) \max(2,c)} \to 4$.
\end{remark}
\begin{remark}
	For the case of symmetric matrices $\bX\in S^p$, the constant $\overline{k}$ can be improved to $\max_{k}[\bB_0]_{k,k} \cdot \nn \bX \bA_0\nn_2$.
\end{remark}

\begin{IEEEproof}
	This proof is based on a large-deviation theory argument. Fix $(k,l) \in \{1,\dots,f\}^2$. Note that $\E[\bT(\bX)]=\bB_*$. First we bound the upper tail probability on the difference $\bT(\bX)-\bB_*$ and then we turn to the lower tail probability. Bounding the upper tail by using Markov's inequality, we have
\begin{align}
	\P & \left([\bT(\bX)]_{k,l}-[\bB_*]_{k,l}>\epsilon \right) \nonumber \\
		&= \P \left(\frac{1}{p}\sum_{i,j=1}^p{\bX_{i,j}[\hat{\bS}_n(j,i)]_{k,l}}-\frac{\tr(\bX\bA_0)}{p} [\bB_0]_{k,l}>\epsilon \right) \nonumber \\
		&= \P \Big(\sum_{m=1}^n \sum_{i,j=1}^p \bX_{i,j} \Big([\bz_m]_{(i-1)f+k}[\bz_m]_{(j-1)f+l} \nonumber \\
		&\qquad - [\bA_0]_{i,j}[\bB_0]_{k,l} \Big) > np \epsilon \Big)  \nonumber \\
		&= \P \Big( \exp\{t \sum_{m=1}^n \sum_{i,j=1}^p \bX_{i,j} \Big([\bz_m]_{(i-1)f+k}[\bz_m]_{(j-1)f+l} \nonumber \\
		&\qquad - [\bA_0]_{i,j}[\bB_0]_{k,l} \Big)\} > \exp\{t np \epsilon\} \Big)  \nonumber \\
		&\leq e^{-tnp\epsilon} \E \Big[\prod_{m=1}^n \exp\Big\{ t \sum_{i,j=1}^p \bX_{i,j} \Big( [\bz_m]_{(i-1)f+k}[\bz_m]_{(j-1)f+l} \nonumber \\
		&\quad -[\bA_0]_{i,j}[\bB_0]_{k,l} \Big) \Big\} \Big] \nonumber \\
		&\leq e^{-tnp\epsilon} \Big( \E\Big[ \exp\Big\{ t \tilde{Y}^{(k,l)} \Big\} \Big] \Big)^n \label{eq2}
\end{align}
where we used the i.i.d. property of the data in (\ref{eq2}) and $\tilde{Y}^{(k,l)}:= \sum_{i,j=1}^p \bX_{i,j}([\bz]_{(i-1)f+k}[\bz]_{(j-1)f+l}-[\bA_0]_{i,j}[\bB_0]_{k,l})$. Define $p^2\times 1$ random vector $\bz^{(k,l)}$ as $[\bz^{(k,l)}]_{(i-1)p+j}:=[\bz]_{(i-1)f+k}[\bz]_{(j-1)f+l} -[\bA_0]_{i,j}[\bB_0]_{k,l}$ for $1\leq i,j\leq p$. Clearly, this random vector is zero mean. The expectation term inside the parentheses in (\ref{eq2}) is the MGF of the random variable $\tilde{Y}^{(k,l)}=\vec(\bX)^T\bz^{(k,l)}$. For notational simplicity, let $\tilde{\phi}_{Y}(t)=\E[e^{tY}]$ denote the MGF of a random vector $Y$. As a result, $\E[e^{t \tilde{Y}^{(k,l)}}]=\tilde{\phi}_{\tilde{Y}^{(k,l)}}(t)$.

Performing a second order Taylor expansion on $\tilde{\phi}_{\tilde{Y}^{(k,l)}}$ about the origin, we obtain:
\begin{equation*}
	\tilde{\phi}_{\tilde{Y}^{(k,l)}}(t) = \tilde{\phi}_{\tilde{Y}^{(k,l)}}(0) + \frac{d\tilde{\phi}_{\tilde{Y}^{(k,l)}}(0)}{dt} t + \frac{1}{2} \frac{d^2 \tilde{\phi}_{\tilde{Y}^{(k,l)}}(\delta t)}{dt^2} t^2
\end{equation*}
for some $\delta \in[0,1]$. Trivially, $ \tilde{\phi}_{\tilde{Y}^{(k,l)}}(0)=1$ and $\frac{d\tilde{\phi}_{\tilde{Y}^{(k,l)}}(0)}{dt}=\E[\vec(\bX)^T \bz^{(k,l)}]=0$. Using the linearity of the expectation operator, we have:
\begin{align*}
	\frac{d^2 \tilde{\phi}_{\tilde{Y}^{(k,l)}}(\delta t)}{dt^2} &= \E[(\tilde{Y}^{(k,l)})^2 e^{t \delta \tilde{Y}^{(k,l)}}] \\
			&= \sum_{m=0}^\infty \frac{(\delta t)^m}{m!} \E[(\vec(\bX)^T \bz^{(k,l)})^{m+2}]
\end{align*}

Using the elementary inequality $1+y \leq e^{y}$ for $y>-1$, and after some algebra, we have:
\begin{equation} \label{eq_temp}
	n\ln(\tilde{\phi}_{\tilde{Y}^{(k,l)}}(t)) \leq \frac{n}{2} t^2 \sum_{m=0}^\infty{T_m(t)}
\end{equation}
where $T_m(t):=\frac{(t\delta)^m}{m!} \E[(\vec(\bX)^T \bz^{(k,l)})^{m+2}]$. Note that
\begin{align}
	t^2 &T_m(t) \leq \frac{t^{m+2}}{m!} \E\Big[\Big(\sum_{i,j=1}^p \bX_{i,j} ([\bz]_{(i-1)f+k}[\bz]_{(j-1)f+l} \nonumber \\
		&\quad -[\bA_0]_{i,j}[\bB_0]_{k,l})\Big)^{m+2}\Big] \nonumber \\
		&=\frac{t^{m+2}}{m!} \sum_{i_1,j_1=1}^p \cdots \sum_{i_{m+2},j_{m+2}=1}^p \bX_{i_1,j_1} \cdots \bX_{i_{m+2},j_{m+2}} \nonumber \\
		& \times \E\Big[ \prod_{\alpha=1}^{m+2} \Big( [\bz]_{(i_\alpha-1)f+k}[\bz]_{(j_\alpha-1)f+l} -[\bA_0]_{i_\alpha,j_\alpha}[\bB_0]_{k,l} \Big) \Big] \nonumber \\
		&\leq \frac{t^{m+2}}{m!} (2m+2)!! p (\max_{1\leq k\leq f} [\bB_0]_{k,k} \nn \bX\nn_2 \nn\bA_0\nn_2)^{m+2} \label{eq3} \\
		&= \frac{(2m+2)!!}{m!} (t \overline{k})^{m+2} p \nonumber
\end{align}
where (\ref{eq3}) follows from Lemma \ref{Isserlis_bound} \footnote{In the symmetric $\bX$ case, this bound can be tightened using $\tr((\bX \bA_0)^{m+2}) \leq p (\nn \bX\bA_0\nn_2)^{m+2}$.}. Also, we defined $\overline{k}=\max_{1\leq k\leq f} [\bB_0]_{k,k} \cdot \nn \bX \nn_2 \nn\bA_0\nn_2$. Summing the result over $m$, and letting $u:=t \overline{k}>0$, $a_m(u):= \frac{(2m+2)!!}{m!} u^m$, $\psi(u):=\sum_{m=0}^\infty a_m(u)$,  we obtain:
\begin{equation} \label{sum_bound}
	t^2 \sum_{m=0}^\infty{T_m(t)} \leq p u^2 \psi(u) \Big|_{u=t \overline{k}}
\end{equation}
By the ratio test \cite{IntroRealAnalysis}, the infinite series $\sum_{m=0}^\infty a_m(u)$ converges if $u<1/2$. To see this, note
\begin{align*}
	\rho &:= \lim_{m\to\infty} \frac{a_{m+1}(u)}{a_m(u)} \\
			&= \lim_{m\to\infty} u \frac{(2m+4)!!}{(2m+2)!!} \frac{m!}{(m+1)!} \\
			&= \lim_{m\to\infty} 2u \frac{1+2/m}{1+1/m} \\
			&= 2u < 1
\end{align*}

Using (\ref{sum_bound}) in (\ref{eq_temp}), and the result in (\ref{eq2}), we obtain the exponential bound:
\begin{align*}
	\P &([\bT(\bX)]_{k,l}-[\bB_*]_{k,l}>\epsilon) \\
		&\leq \exp\Big\{-tnp\epsilon + \frac{np(t \overline{k})^2}{2}  \psi(t \overline{k})\Big\}
\end{align*}
Let $t<\frac{1}{(2+\tau) \overline{k}}$ and $\epsilon<\frac{1}{2+\tau} \psi(\frac{1}{2+\tau}) \overline{k}<\infty$. By the monotonicity of $\psi(\cdot)$, we have:
\begin{equation}
	\P([\bT(\bX)]_{k,l}-[\bB_*]_{k,l}>\epsilon) \leq \exp \Big\{-tnp\epsilon + \frac{np t^2 \overline{k}^2}{2}  \psi(\frac{1}{2+\tau}) \Big\} \label{exp_bound}
\end{equation}
Optimizing (\ref{exp_bound}) over $t$, we obtain $t^*=\frac{\epsilon}{\overline{k}^2 \psi(\frac{1}{2+\tau})}$. Clearly, $t^*<\frac{1}{(2+\tau) \overline{k}}$. Plugging this into (\ref{exp_bound}), we obtain:
\begin{equation*}
	\P([\bT(\bX)]_{k,l}-[\bB_*]_{k,l}>\epsilon) \leq \exp \Big\{-\frac{np\epsilon^2}{2 \overline{k}^2 \psi(\frac{1}{2+\tau})} \Big\}
\end{equation*}
Define $C:=\frac{1}{2 \overline{k}^2 \psi(\frac{1}{2+\tau})}$. Since $\psi(\frac{1}{2+\tau})<\infty$, $C>0$. Thus, for all $\epsilon < \frac{1}{2+\tau} \psi(\frac{1}{2+\tau}) \overline{k}$, we have
\begin{equation} \label{upper_tail_2}
	P([\bT(\bX)]_{k,l}-[\bB_*]_{k,l}>\epsilon) \leq e^{-np\epsilon^2 C}
\end{equation}
where $C>0$ is independent of $n,p,f$.

 Next, we bound the lower tail:
	\begin{align*}
		\P &([\bT(\bX)]_{k,l}-\E[[\bT(\bX)]_{k,l}]<-\epsilon) \\
			&= \P \Big(\sum_{m=1}^n \sum_{i,j=1}^p -\bX_{i,j}([\bz_m]_{(j-1)f+k}[\bz_m]_{(i-1)f+l} \\
			&\quad -[\bA_0]_{i,j}[\bB_0]_{k,l}) > np\epsilon \Big) \\
			&\leq e^{-tnp\epsilon} \left( \tilde{\phi}_{\tilde{Y}^{(k,l)}}(-t) \right)^n
	\end{align*}
	where $\tilde{\phi}_{\tilde{Y}^{(k,l)}}$ is the MGF of $\tilde{Y}^{(k,l)}$. Performing a second order Taylor expansion as before, we have:
	\begin{align*}
		\tilde{\phi}_{\tilde{Y}^{(k,l)}}(-t) &= \tilde{\phi}_{\tilde{Y}^{(k,l)}}(0) - \frac{d\tilde{\phi}_{\tilde{Y}^{(k,l)}}(0)}{dt} t + \frac{1}{2} \frac{d^2 \tilde{\phi}_{\tilde{Y}^{(k,l)}}(\delta t)}{dt^2} t^2 \\
		&= 1 + \frac{t^2}{2} \sum_{m=0}^\infty T_m'(t)
	\end{align*}
	where $T_m'(t):= \frac{(-t\delta)^m}{m!} \E[(<\vec(\bX),\bz^{(k,l)}>)^{m+2}] = (-1)^m T_m(t) \leq T_m(t)$ and $\delta\in[0,1]$. Proceeding similarly as above, it can be shown that for all $\epsilon<\frac{1}{2+\tau} \psi(\frac{1}{2+\tau}) \overline{k}$:
	\begin{equation} \label{lower_tail_2}
			\P([\bT(\bX)]_{k,l}-\E[[\bT(\bX)]_{k,l}]<-\epsilon) \leq e^{-np\epsilon^2 C}
	\end{equation}
	where $C$ was defined as before. From (\ref{upper_tail_2}) and (\ref{lower_tail_2}), we conclude that for all $\epsilon<\frac{1}{2+\tau} \psi(\frac{1}{2+\tau}) \overline{k}$:
	\begin{align*}
			\P &(|[\bT(\bX)]_{k,l}-\E[[\bT(\bX)]_{k,l}]|>\epsilon) \\
			&\leq \P([\bT(\bX)]_{k,l}-\E[[\bT(\bX)]_{k,l}]>\epsilon) \\
			&\quad + \P([\bT(\bX)]_{k,l}-\E[[\bT(\bX)]_{k,l}]<-\epsilon) \\
			&\leq 2 e^{-np\epsilon^2 C}
	\end{align*}
	The union bound over $(k,l)\in \{1,\dots,f\}^2$ completes the proof. Let us rewrite this. If $\frac{4 \max(2,c) \log(\max(f,n)) (2+\tau)^2}{\psi(\frac{1}{2+\tau})}<np$, then with probability $1-\frac{2}{\max(f,n)^c}$,
	\begin{align*}
		|\bT(\bX)-\E[\bT(\bX)]|_\infty &\leq \overline{k} \cdot \sqrt{2 \psi(\frac{1}{2+\tau})} \sqrt{\frac{\log((2f^2)/(2/\max(f,n)^c))}{np}} \\
			&\leq \overline{k} \cdot 2 \sqrt{\psi(\frac{1}{2+\tau}) \max(2,c)} \sqrt{\frac{\log \max(f,n)}{np}}
	\end{align*}
\end{IEEEproof}

\section{Proposition \ref{prop: Glasso_optimal_rate}}
\begin{proposition} \label{prop: Glasso_optimal_rate}
	Let $\bS_{p,f,n}$ be a $d'\times d'$ (where $d'=p$ or $d'=f$) random matrix such that with probability $1-\frac{2}{n^2}$, $|\bS_{p,f,n}-\bSigma_*|_\infty \leq r_{p,f,n}$. Assume $\bSigma_*\in S_{++}^{d'}$ has uniformly bounded spectrum as $p,f\to\infty$ (analog to Assumption 1). Choose $\lambda_{p,f,n} = c \cdot r_{p,f,n}$ for some absolute constant $c>0$. Consider the Glasso operator $\bG(\cdot,\cdot)$ defined in (\ref{G_operator}). Let $s=s_{\bTheta_*}$ be the sparsity parameter associated with $\bTheta_*:=\bSigma_*^{-1}$. Assume $\sqrt{d'+s} \cdot r_{p,f,n}=o(1)$. Then, with probability $1-\frac{2}{n^2}$,
	\begin{equation*}
		\nn \bG(\bS_{p,f,n},\lambda_{p,f,n})-\bTheta_* \nn_F \leq \frac{2\sqrt{2} (1+c)}{\lambda_{min}(\bSigma_*)^2} \sqrt{d'+s} \cdot r_{p,f,n}
	\end{equation*}
	as $n\to\infty$.
\end{proposition}
\begin{IEEEproof}
	The proof follows from a slight modification of Thm. 1 in \cite{Rothman}, or Thm. 3 in \cite{TimeVaryingGraphs}. This modification is due to the different $r_{p,f,n}$.
\end{IEEEproof}

\section{Proof of Theorem \ref{thm: FF_optimal_rate}}
\begin{IEEEproof}
As in the proof of Thm. 1 in \cite{EstCovMatKron}, let $\bB_* = \frac{\tr(\bA_0\bA_{init}^{-1})}{p} \bB_0$ and $\bA_* = (\frac{\tr(\bA_0\bA_{init}^{-1})}{p})^{-1} \bA_0$. Note that Assumption 1 implies that $\nn \bB_* \nn_2 = \Theta(1)$ and $\nn \bA_* \nn_2=\Theta(1)$ as $p,f\to\infty$. For conciseness, the statement ``with probability $1-c n^{-2}$ (where $c>0$ is a constant independent of $p,f,n$)'' will be abbreviated as ``w.h.p.''-i.e., with high probability.

For concreteness, we first present the result for $k=2$ iterations. Then, we generalize the analysis to all finite flip-flop iterations by induction. The growth assumptions in the theorem imply
\begin{equation} \label{FF_sufficient_condition}
	\max \left\{p,f, \frac{f^2}{p}, \left(\frac{\sqrt{pf} + f\sqrt{\frac{f}{p}} + p \sqrt{\frac{p}{f}}}{p+f}\right)^2 \right\} \log M \leq C'n
\end{equation}
for some constant $C'>0$ large enough \footnote{This constant is independent of $p,f,n$, but may depend on the constants in Assumption \ref{assumption_posdef_unif}.}. In fact, the growth assumption in the theorem statement can be relaxed to (\ref{FF_sufficient_condition}).

As in the proof of Thm. 1 in \cite{EstCovMatKron}, we vectorize the operations (\ref{A_update}) and (\ref{B_update}):
\begin{align*}
	\vec(\hat{\bA}(\bB)) &= \frac{1}{f} \hat{\bR}_A \vec(\bB^{-1}) \\
	\vec(\hat{\bB}(\bA)) &= \frac{1}{p} \hat{\bR}_B \vec(\bA^{-1})
\end{align*}
where $\hat{\bR}_A$ and $\hat{\bR}_B$ are permuted versions of the sample covariance matrix \cite{EstCovMatKron}.

Define intermediate error matrices:
\begin{align*}
	\tilde{\bB}^0 &= \hat{\bB}(\bA_{init}) - \bB_* \\
	\tilde{\bA}^1 &= \hat{\bA}(\hat{\bB}(\bA_{init})) - \bA_*
\end{align*}
Define $\bY_*=\bB_*^{-1}$ and $\bX_*=\bA_*^{-1}$. Also, define:
\begin{align*}
	\bY_1 &= \hat{\bB}(\bA_{init})^{-1} \\
	\bX_2 &= \hat{\bA}(\hat{\bB}(\bA_{init}))^{-1}
\end{align*}
These inverses exist if $n\geq \max(\frac{p}{f},\frac{f}{p}) + 1$ (see \cite{LuZimmerman}). Define the error $\tilde{\bSigma}_{FF}(k)=\bSigma_{FF}(k) - \bSigma_0$ for $k \geq 2$. For notational simplicity, let $\bB_0^{max}:=\max_{k}[\bB_0]_{k,k}$ and $\bA_0^{max}:=\max_{i}[\bA_0]_{i,i}$, $\psi_{\tau} := \psi(\frac{1}{2+\tau})$, where $\psi(\cdot)$ is defined in Lemma \ref{lemma: large_dev_optimal}.

Lemma \ref{lemma: large_dev_optimal} implies that for
\begin{equation} \label{cond_0}
	n> \frac{8(2+\tau)^2}{\psi_{\tau}} \log M
\end{equation}
then with probability $1-2 n^{-2}$, we have:
\begin{equation} \label{B_0_Frob1}
	\nn \tilde{\bB}^0 \nn_F \leq C_0 fp^{-1/2} \sqrt{ \frac{\log M}{n}}
\end{equation}
where $C_0=2\sqrt{2\psi_\tau} \bB_0^{max} \nn\bA_{init}^{-1} \bA_0\nn_2$.

Let $\epsilon'>1$. Note that from (\ref{B_0_Frob1}), for
\begin{equation} \label{cond_1}
	n\geq (\epsilon' C_0)^2 f^2 p^{-1} \log M
\end{equation}
with probability $1-2 n^{-2}$,
\begin{align*}
	\lambda_{min} &(\hat{\bB}(\bA_{init})) = \lambda_{min}(\tilde{\bB}^0 + \bB_*) \geq \lambda_{min}(\bB_*) - \nn \tilde{\bB}^0 \nn_2 \\
		&\geq \lambda_{min}(\bB_*) - \nn \tilde{\bB}^0 \nn_F \geq \left( 1-\frac{1}{\epsilon'} \right) \lambda_{min}(\bB_*) > 0
\end{align*}
Thus, letting $\bDelta_Y^1=\bY_1-\bY_*$, w.h.p.,
\begin{align}
	\nn & \bDelta_Y^1 \nn_F = \nn \bY_1 (\hat{\bB}(\bA_{init})-\bB_*) \bY_* \nn_F \nonumber \\ 
		&\leq \nn \bY_1 \nn_2 \nn \bY_* \nn_2 \nn \tilde{\bB}^0 \nn_F = \frac{\nn \tilde{\bB}^0 \nn_F}{ \lambda_{min}(\bB_*) \lambda_{min}(\hat{\bB}(\bA_{init})) } \nonumber \\
		&\leq C_0 \left( 1-\frac{1}{\epsilon'} \right)^{-1} \nn\bY_*\nn_2^2 fp^{-1/2} \sqrt{\frac{\log M}{n}} \label{bound_Frob_spec}
\end{align}


Expanding $\tilde{\bA}^1$:
\begin{align}
	\vec(\tilde{\bA}^1) &= \frac{1}{f} \hat{\bR}_A \vec(\bY_1) - \vec(\bA_*) \nonumber \\
		&= \frac{\tr(\bB_0 \bDelta_Y^1)}{f} \vec(\bA_0) + \vec(\hat{\bA}(\bB_*)-\bA_*) \nonumber \\
		&\quad + \frac{1}{f} \tilde{\bR}_A \vec(\bDelta_Y^1) \label{A_1_error2}
\end{align}
where we used $\bR_A = \vec(\bA_0)\vec(\bB_0^T)^T$ (see Eq. (91) from \cite{EstCovMatKron}). Using the triangle inequality in (\ref{A_1_error2}), the Cauchy-Schwarz inequality, and standard matrix norm bounds:
\begin{align}
	\nn \tilde{\bA}^1 \nn_F &\leq \underbrace{\sqrt{\frac{p}{f}} \nn \bSigma_0\nn_2 \nn \bDelta_Y^1 \nn_F}_{T_1} + \underbrace{p |\hat{\bA}(\bB_*)-\bA_*|_\infty}_{T_2} \nonumber \\
		&\quad + \underbrace{\frac{p}{f} \nn \tilde{\bR}_A \vec(\bDelta_Y^1) \nn_\infty}_{T_3} \nonumber
\end{align}
We note upon expanding:
\begin{equation*}
	\frac{1}{f} \nn \tilde{\bR}_A \vec(\bDelta_Y^1) \nn_\infty = \left|\frac{1}{f}\sum_{k,l=1}^f [\bDelta_Y^1]_{k,l} \bar{\hat{\bS}}_n(k,l) - \frac{\tr(\bB_0\bDelta_Y^1)}{f} \bA_0\right|_\infty
\end{equation*}
From (\ref{bound_Frob_spec}), there exists $c>0$ such that:
\begin{equation*}
	\P\left( T_1 \geq C_1 f^{1/2} \sqrt{\frac{\log M}{n}} \right) \leq c n^{-2}
\end{equation*}
where $C_1=\nn \bSigma_0\nn_2 C_0 (1-1/\epsilon')^{-1} \nn \bY_* \nn_2^2$ is an absolute constant. Lemma \ref{lemma: large_dev_optimal} implies:
\begin{equation*}
	\P\left( T_2 \geq C_2 f^{-1/2} \sqrt{\frac{\log M}{n}} \right) \leq 2 n^{-2}
\end{equation*}
where $C_2 = 2\sqrt{2 \psi_\tau} A_0^{max} \nn \bY_* \bB_0\nn_2$ is an absolute constant. To bound $T_3$, we define the following events:
\begin{align*}
	E_0 &= \left\{ \nn \bDelta_Y^1 \nn_F \leq \frac{C_1}{\nn \bSigma_0 \nn_2} fp^{-1/2} \sqrt{\frac{\log M}{n}} \right\} \\
	E_1 &= \left\{ \Big|\frac{1}{f}\sum_{k,l=1}^f [\bDelta_Y^1]_{k,l} \bar{\hat{\bS}}_n(k,l) - \frac{\tr(\bB_0\bDelta_Y^1)}{f} \bA_0 \Big|_\infty \leq 2\sqrt{2 \psi_\tau} A_0^{max} \nn\bDelta_Y^1 \nn_F \nn\bB_0 \nn_2 \sqrt{\frac{\log M}{nf}} \right\} \\
	E_2 &= \left\{ T_3  \leq C_3 \sqrt{pf} \sqrt{\frac{\log M}{n}} \right\}
\end{align*}
where $C_3=2\sqrt{2\psi_\tau} A_0^{max} \nn\bB_0\nn_2 C_0 (1-1/\epsilon')^{-1} \nn\bY_*\nn_2^2$ is an absolute constant. From (\ref{bound_Frob_spec}), it follows that $\P\left( E_0 \right) \geq 1 - cn^{-2}$ and from Lemma (\ref{lemma: large_dev_optimal}), it follows that $\P\left( E_1 | E_0 \right) \geq 1 - 2n^{-2}$. As a result, we have $\P(E_2) \geq \P(E_1 \cap E_0) = \P(E_1|E_0) \P(E_0) \geq 1-(c+2)n^{-2}$. Putting it together with the union bound, we have:
\begin{align}
	\P\Bigg( & \nn \tilde{\bA}^1 \nn_F \geq (C_1 f^{1/2} + C_2 pf^{-1/2})\sqrt{\frac{\log M}{n}} + C_3 \sqrt{pf} \frac{\log M}{n}  \Bigg) \nonumber \\
		&\leq \P\left(T_1 \geq \frac{C_1}{3} f^{1/2} \sqrt{\frac{\log M}{n}} \right) + \P\left( T_2 \geq \frac{C_2}{3} pf^{-1/2} \sqrt{\frac{\log M}{n}} \right) \nonumber \\
		&\quad + \P\left(T_3 \geq \frac{C_3}{3} \sqrt{pf} \frac{\log M}{n} \right) \nonumber \\
		&\leq c' n^{-2}	\label{tilde_A_1_Frob_bound}
\end{align}
for some $c'>0$ absolute constant.



Let $c_1>0$. For
\begin{equation} \label{cond_3}
	n \geq \left( \frac{C_3}{c_1 \max(C_1,C_2)} \right)^2 \frac{pf}{(f^{1/2}+pf^{-1/2})^2} \log M
\end{equation}
then, from (\ref{tilde_A_1_Frob_bound}), we have w.h.p.,
\begin{equation} \label{tilde_A_1_Frob2}
	\nn \tilde{\bA}^1 \nn_F \leq \max(C_1,C_2) (1+c_1) (\sqrt{f}+pf^{-1/2}) \sqrt{\frac{\log M}{n}}
\end{equation}

Using properties of the Kronecker product:
\begin{align}
	\tilde{\bSigma}_{FF}(2) &= \tilde{\bA}^1 \otimes \bB_* + \bA_* \otimes \tilde{\bB}^0 \nonumber \\
		&\quad + \tilde{\bA}^1 \otimes \tilde{\bB}^0 \label{error_FF_2iter}
\end{align}
From (\ref{B_0_Frob1}),(\ref{tilde_A_1_Frob2}), (\ref{error_FF_2iter}), under conditions (\ref{cond_0}),(\ref{cond_1}), and (\ref{cond_3}), w.h.p.,
\begin{align}
	\nn &\tilde{\bSigma}_{FF}(2) \nn_F \leq \nn \tilde{\bA}_1 \nn_F \nn \bB_*\nn_F \nonumber \\
		&\quad + \nn \bA_*\nn_F \nn \tilde{\bB}^0 \nn_F + \nn \tilde{\bA}^1\nn_F \nn \tilde{\bB}^0\nn_F \nonumber \\
		&\leq \tilde{C}_3 (p+2f) \sqrt{\frac{\log M}{n}} + \tilde{C}_4 (f\sqrt{f/p}+\sqrt{pf}) \frac{\log M}{n} \label{tilde_R_FF_2}
\end{align}
where $\tilde{C}_3=\max(\nn\bB_*\nn_2 \max(C_1,C_2) (1+c_1),C_0 \nn \bA_*\nn_2)$ and $\tilde{C}_4=C_0 \max(C_1,C_2) (1+c_1)$ are constants.

Let $c_2>0$. For
\begin{equation*}
	n \geq (\frac{\tilde{C}_4}{\tilde{C}_3 c_2})^2 \frac{(f\sqrt{f/p}+\sqrt{pf})^2}{(p+2f)^2} \log M
\end{equation*}
then, from (\ref{tilde_R_FF_2}) w.h.p.,
\begin{equation*}
	\nn \tilde{\bSigma}_{FF}(2)\nn_F \leq \tilde{C}_3(1+c_2) (p+2f) \sqrt{\frac{\log M}{n}}
\end{equation*}

The proof for $k=2$ iterations is complete. Using a simple induction, it follows that the rate (\ref{FF_rate_2}) holds for all $k$ finite.

Next, we show that the convergence rate in the precision matrix Frobenius error is on the same order as the covariance matrix error. Let $\bTheta_{FF}(2):=\bSigma_{FF}(2)^{-1}$. From (\ref{tilde_A_1_Frob2}), for
\begin{equation*}
	n> (\epsilon' \nn\bX_*\nn_2 \max(C_1,C_2) (1+c_1))^2 (\sqrt{f}+pf^{-1/2})^2 \log M
\end{equation*}
then, letting $\bDelta_X^2=\bX_2-\bX_*$, we have w.h.p.,
\begin{align}
	\nn \bDelta_X^2 \nn_F &\leq \left( 1-\frac{1}{\epsilon'} \right)^{-1} \nn\bX_*\nn_2^2 \tilde{C}_1(1+c_1) \nonumber \\
		&\quad \times (\sqrt{f}+pf^{-1/2}) \sqrt{\frac{\log M}{n}} \label{X_2_Frob_error}
\end{align}
Using (\ref{bound_Frob_spec}) and (\ref{X_2_Frob_error}), we have w.h.p.,
\begin{align}
	\nn & \bTheta_{FF}(2)-\bTheta_0 \nn_F \leq \nn \bDelta_X^2 \nn_F \nn \bY_*\nn_F \nonumber \\
		&\quad + \nn \bDelta_Y^1 \nn_F \nn \bX_*\nn_F + \nn \bDelta_X^2 \nn_F \nn \bDelta_Y^1 \nn_F \nonumber \\
		&\leq \tilde{D}_1 (2f+p) \sqrt{\frac{\log M}{n}} + \tilde{D}_2 (f\sqrt{\frac{f}{p}}+\sqrt{pf}) \frac{\log M}{n} \label{bound_Theta}
\end{align}
where $\tilde{D}_1$ and $\tilde{D}_2$ are constants.

For
\begin{equation*}
	n>(\frac{\tilde{D}_2}{\tilde{D}_1 d'})^2 (\frac{f\sqrt{f/p}+\sqrt{pf}}{2f+p})^2 \log M
\end{equation*}
the bound (\ref{bound_Theta}) becomes w.h.p.,
\begin{equation*}
	\nn \bTheta_{FF}(2)-\bTheta_0 \nn_F \leq \tilde{D}_1 (1+d') (2f+p) \sqrt{\frac{\log M}{n}}
\end{equation*}
Thus, the same rate $O_P\left( \sqrt{\frac{(p^2+f^2)\log M}{n}} \right)$ holds for the precision matrix Frobenius error.

%
%
%

\end{IEEEproof}

\section{Proof of Theorem \ref{thm: KGL_optimal_rate}}
\begin{IEEEproof}
We show that the first iteration of the KGL algorithm yields a fast statistical convergence rate of $O_P\left( \sqrt{\frac{(p+f)\log M}{n}} \right)$ by appropriately adjusting the regularization parameters. A simple induction finishes the proof. Adopt the notation from the proof of Thm. \ref{thm: FF_optimal_rate}.

Lemma \ref{lemma: large_dev_optimal} implies that for
\begin{equation} \label{condition_0}
	n \geq \frac{8(2+\tau)^2}{\psi_{\tau}} \log M
\end{equation}
then with probability $1-2n^{-2}$,
\begin{equation} \label{tilde_B_0_rate}
	|\tilde{\bB}^0|_\infty \leq C_0 p^{-1/2} \sqrt{\frac{\log M}{n}}
\end{equation}
where $\tilde{\bB}^0=\hat{\bB}(\bA_{init})-\bB_*$. From Proposition \ref{prop: Glasso_optimal_rate} and (\ref{tilde_B_0_rate}), we obtain w.h.p.,
\begin{align}
	\nn & \bY_1-\bY_* \nn_F \leq 2\sqrt{2} (1+c_y) \sqrt{1+c_{Y_0}} \nn \bY_*\nn_2^2 \nonumber \\
		&\times C_0 \sqrt{\frac{f\log M}{np}} \label{Y_Frob_error}
\end{align}
where we also used $s_{Y_0}\leq c_{Y_0} f$ and $\bY_1:=\bG(\hat{\bB}(\bA_{init}),\lambda_Y^{(1)})=\bB_1^{-1}$. Note that $fp^{-1}\log M=o(n)$ was used here. Let $\bDelta_Y^1=\bY_1-\bY_*$.

Let $\grave{\bA}^1:=\hat{\bA}(\bB_1) - \bA_*$. Then, we have
\begin{align}
	\vec(\grave{\bA}^1) &= \frac{1}{f} \hat{\bR}_A \vec(\bY_1) - \vec(\bA_*) \nonumber \\
		&= \frac{\tr(\bB_0 \bDelta_Y^1)}{f} \vec(\bA_0) + \vec(\hat{\bA}(\bB_*)-\bA_*) \nonumber \\
		&\quad + \frac{1}{f} \tilde{\bR}_A \vec(\bDelta_Y^1) \label{grave_A_1}
\end{align}
where we used $\bR_A=\vec(\bA_0)\vec(\bB_0^T)^T$ (see Eq. (91) in \cite{EstCovMatKron}).


From (\ref{grave_A_1}), applying the triangle inequality and using the Cauchy-Schwarz inequality:
\begin{align}
	|\grave{\bA}^1|_\infty 
		&\leq \underbrace{\frac{\sqrt{f} \nn \bB_0\nn_2 \nn \bDelta_Y^1 \nn_F}{f} |\bA_0|_\infty}_{T_1} + \underbrace{|\hat{\bA}(\bB_*)-\bA_*|_\infty}_{T_2} \nonumber \\
		&\quad + \underbrace{\frac{1}{f}\nn \tilde{\bR}_A\vec(\bDelta_Y^1) \nn_\infty}_{T_3} \label{grave_A_1_bound} \\
\end{align}
Let $\tilde{C}_0=C_0 2\sqrt{2} (1+c_y) \sqrt{1+c_{Y_0}} \nn \bY_* \nn_2^2$ and $\bar{C}_1 = \tilde{C}_0 |\bA_0|_\infty \nn \bB_0\nn_2$. The bound (\ref{Y_Frob_error}) implies
\begin{equation*}
	\P\left( T_1 \geq \bar{C}_1 \sqrt{\frac{\log M}{np}} \right) \leq c n^{-2}
\end{equation*}
for some $c>0$. Let $\bar{C}_2=2\sqrt{2\psi_\tau} A_0^{max} \nn \bY_* \bB_0\nn_2$. Lemma \ref{lemma: large_dev_optimal} implies
\begin{equation*}
	\P\left( T_2 \geq \bar{C}_2 \sqrt{\frac{\log M}{nf}} \right) \leq 2 n^{-2}
\end{equation*}
Let $\bar{C}_3=\tilde{C}_0 2\sqrt{2\psi_\tau} A_0^{max} \nn\bB_0\nn_2$. To bound $T_3$, we use the same technique as in the proof of Thm. \ref{thm: FF_optimal_rate}. Define the events:
\begin{align*}
	E_0 &= \left\{ \nn \bDelta_Y^1\nn_F \leq \tilde{C}_0 \sqrt{\frac{f \log M}{ np }} \right\} \\
	E_1 &= \left\{ \frac{1}{f} \nn \tilde{\bR}_A \vec(\bDelta_Y^1) \nn_\infty \leq 2\sqrt{2\psi_\tau} A_0^{max} \nn\bB_0\nn_2 \nn\bDelta_Y^1\nn_F \sqrt{\frac{\log M}{nf}} \right\} \\
	E_2 &= \left\{ T_3 \leq \bar{C}_3 \frac{1}{\sqrt{p}} \frac{\log M}{n} \right\} 
\end{align*}
From (\ref{Y_Frob_error}), we have $\P(E_0)\geq 1-cn^{-2}$ and from Lemma \ref{lemma: large_dev_optimal} we have $\P(E_1|E_0) \geq 1- 2n^{-2}$. Thus, $\P(E_2) \geq \P(E_1|E_0) \P(E_0) \geq 1-c'n^{-2}$.

Using (\ref{grave_A_1_bound}) and the union bound:
\begin{align*}
	\P &\left( |\grave{\bA}^1|_\infty \geq (\frac{\bar{C}_1}{\sqrt{p}}+\frac{\bar{C}_2}{\sqrt{f}}) \sqrt{\frac{\log M}{n}} + \frac{\bar{C}_3}{\sqrt{p}} \frac{\log M}{n} \right) \\
		&\leq \P\left( T_1 \geq \frac{\bar{C}_1}{3 \sqrt{p}} \sqrt{\frac{\log M}{n}} \right) + \P\left(T_2 \geq \frac{\bar{C}_2}{3\sqrt{f}} \sqrt{\frac{\log M}{n}} \right) \\
		&\quad + \P\left( T_3 \geq \frac{\bar{C}_3}{3 \sqrt{p}} \frac{\log M}{n} \right) \\
		&\leq c'' n^{-2}
\end{align*}
for some $c''>0$. Thus, for $n\geq (\frac{\bar{C}_3}{\bar{C}_1 c_1})^2 \log M$, $c_1>0$, we have w.h.p.,
\begin{equation} \label{grave_A_1_ub}
	|\grave{\bA}^1|_\infty \leq \max(\bar{C}_1,\bar{C}_2) (1+c_1) \left(\frac{1}{\sqrt{p}}+\frac{1}{\sqrt{f}}\right) \sqrt{\frac{\log M}{n}}
\end{equation}

Let $\bDelta_X^1=\bX_1-\bX_*$. From Proposition \ref{prop: Glasso_optimal_rate} and (\ref{grave_A_1_ub}), we obtain w.h.p.:
\begin{align}
	\nn & \bDelta_X^1 \nn_F \leq 2\sqrt{2}(1+c_x)\sqrt{1+c_{X_0}} \nn\bX_*\nn_2^2 \max(\bar{C}_1,\bar{C}_2) (1+c_1) \nonumber \\
		&\quad \times \left(1 + \sqrt{\frac{p}{f}} \right) \sqrt{\frac{\log M}{n}} \label{X_Frob_error}
\end{align}
where we used $s_{X_0}\leq c_{X_0} p$ and $\bX_1:=\bG(\hat{\bA}(\bB_1), \lambda_X^{(1)})$, $\bX_*:=\bA_*^{-1}$. Note that $(1+\sqrt{p/f})^2\log M=o(n)$ was used here.

Finally, using (\ref{Y_Frob_error}) and (\ref{X_Frob_error}), we obtain w.h.p.:
\begin{align}
	& \nn \bTheta_{KGL}(2)-\bTheta_0 \nn_F = \nn \bX_1 \otimes \bY_1 - \bX_* \otimes \bY_* \nn_F \nonumber \\
		&\leq \nn \bDelta_Y^1 \nn_F \sqrt{p} \nn \bX_*\nn_2 + \nn \bDelta_X^1 \nn_F \sqrt{f} \nn \bY_*\nn_2 \nonumber \\
		&\quad + \nn \bDelta_Y^1 \nn_F \nn \bDelta_X^1 \nn_F \nonumber \\
		&\leq \bar{C'}_3 (2\sqrt{f}+\sqrt{p}) \sqrt{\frac{\log M}{n}} + \bar{C'}_4 (1+\sqrt{\frac{f}{p}}) \frac{\log M}{n} \label{Theta_bound}
\end{align}
where $\bar{C'}_3$ and $\bar{C'}_4$ are constants \cite{TsiligkaridisTSP}.
For
\begin{equation*}
	n>(\frac{\bar{C'}_4}{\bar{C'}_3 \bar{c} })^2 \left(\frac{1+\sqrt{f/p}}{2\sqrt{f}+\sqrt{p}}\right)^2 \log M
\end{equation*}
the bound (\ref{Theta_bound}) further becomes:
\begin{equation*}
	\nn \bTheta_{KGL}(2)-\bTheta_0 \nn_F \leq \bar{C'}_3 (1+\bar{c}) (2\sqrt{f}+\sqrt{p}) \sqrt{\frac{\log M}{n}}
\end{equation*}

Note that $\nn \bTheta_{KGL}(2)-\bTheta_0 \nn_F^2 = O_P\left( \frac{ (p+f+\sqrt{pf}) \log M}{n} \right) =  O_P\left( \frac{ (p+f) \log M}{n} \right)$ as $p,f,n\to\infty$. This concludes the first part of the proof. The rest of the proof follows by similar bounding arguments coupled with induction. The rate remains the same as the number of iterations increases, but the constant on front may change.

Next, we show that the convergence rate in the covariance matrix Frobenius error is on the same order as the inverse. From (\ref{Y_Frob_error}), for
\begin{equation*}
	n> (\epsilon' \tilde{C}_0 \nn \bY_*\nn_2)^2 fp^{-1} \log M
\end{equation*}
we have w.h.p. $\lambda_{min}(\bY_1) \geq \lambda_{min}(\bY_*)-\nn \bY_1-\bY_* \nn_F\geq (1-\frac{1}{\epsilon'}) \lambda_{min}(\bY_*)$, which in turn implies w.h.p.,
\begin{align}
	 \nn \bDelta_B^1 \nn_F &= \nn \bB_1-\bB_* \nn_F \leq \underbrace{(1-1/\epsilon')^{-1} \tilde{C}_0 \nn\bB_*\nn_2^2}_{\bar{C}_B^1} \nonumber \\
		&\quad \times \sqrt{\frac{f}{p}} \sqrt{\frac{\log M}{n}} \label{B1_Frob_error}
\end{align}
\footnote{Here, $\bB_1 = \bY_1^{-1}$ exists since $\bY_1$ is positive definite (see (\ref{G_operator})).}
Using a similar argument, from (\ref{X_Frob_error}), for $n \geq C' (1+\sqrt{\frac{p}{f}})^2 \log M$ (for some constant $C'$)
we have w.h.p.,
\begin{align}
	\nn\bDelta_A^1 \nn_F &= \nn \bA_1-\bA_* \nn_F \leq \underbrace{(1-1/\epsilon')^{-1} \nn \bA_* \nn_2^2 \bar{C}_X^1}_{\bar{C}_A^1} \nonumber \\
		&\quad \times \left(1+\sqrt{\frac{p}{f}}\right) \sqrt{\frac{\log M}{n}} \label{A1_Frob_error}
\end{align}
where $\bA_1=\bX_1^{-1}$.

Let $\bSigma_{KGL}(2):=\bTheta_{KGL}(2)^{-1}=\bA_1\otimes \bB_1$. Then, w.h.p.,
\begin{align}
	\nn & \bSigma_{KGL}(2)-\bSigma_0 \nn_F \leq \nn \bDelta_A^1 \nn_F \nn \bB_*\nn_F \nonumber \\
		&\quad + \nn \bDelta_B^1 \nn_F \nn \bA_*\nn_F + \nn\bDelta_A^1\nn_F \nn\bDelta_B^1\nn_F \nonumber \\
		&\leq \bar{D}_1 (2\sqrt{f}+\sqrt{p}) \sqrt{\frac{\log M}{n}} + \bar{D}_2 (1+\sqrt{\frac{f}{p}}) \frac{\log M}{n} \label{Sigma_KGL_error}
\end{align}
where $\bar{D}_1$ and $\bar{D}_2$ are constants \cite{TsiligkaridisTSP}.
For
\begin{equation*}
	n>(\frac{\bar{D}_2}{\bar{D}_1 d})^2 \left(\frac{1+\sqrt{\frac{f}{p}}}{2\sqrt{f}+\sqrt{p}}\right)^2 \log M
\end{equation*}
then (\ref{Sigma_KGL_error}) implies w.h.p.,
\begin{equation*}
	\nn \bSigma_{KGL}(2)-\bSigma_0 \nn_F \leq \bar{D}_1 (1+d) (2\sqrt{f}+\sqrt{p}) \sqrt{\frac{\log M}{n}}
\end{equation*}
Thus, the same rate $O_P\left( \sqrt{\frac{(p+f) \log M}{n}} \right)$ holds for the error in the covariance matrix.

\end{IEEEproof}

\bibliographystyle{IEEEtran}
\bibliography{myRefs}

\end{document}